\documentstyle[a4,12pt,epsfig,rotating,here,amssymb,graphics]{article}

\begin{document}

\begin{titlepage}

\pagenumbering{arabic}
\vspace*{-1.5cm}
\begin{tabular*}{14.cm}{l@{\extracolsep{\fill}}r}
& LUNFD6/(NFFL-7175) 1999
\\
& 
27 October 1999
\\ \hline
\end{tabular*}
\vspace*{2.cm}

\large
\centerline {\bf BEAUTY'99 Conference Summary}
\normalsize
 
\vskip 2.0cm
\centerline {Paula Eerola~\footnote{paula.eerola@quark.lu.se}}
\centerline {\it Lund University, Lund, Sweden}
\vskip 4.0cm
 
\centerline {\bf Abstract}
\vskip 1.0cm
Investigations of B hadrons are expected to
break new ground in measuring CP-violation effects. This series of 
BEAUTY conferences, originating from the 1993 conference in Liblice,
has contributed significantly in developing ideas of CP-violation
measurements using B hadrons and formulating and comparing critically
the B-physics experiments. In the '99 conference in Bled we saw 
the ripening of the field and the first fruit emerging --
Tevatron have produced beautiful B-physics results and more are expected
to come with the next run, while the B-physics experiments at DESY, 
SLAC and KEK
are starting their operation. The longer-term projects at LHC and Tevatron
have taken their shape and detailed prototyping work is going on. Meanwhile,
on the phenomenological side, there has been impressive theoretical progress 
in understanding deeper the `standard'
measurements and proposing new signatures. 
In this summary, I will highlight the status of the field
as presented in the conference, concentrating on signatures, experiments,
and R\&D programmes. 

\vfill
\end{titlepage}

\newpage

\section{Introduction}

CP violation is a key phenomenon in elementary particle physics for
completing our present understanding of particles and interactions,
formulated by the Standard Model of electroweak interactions, and searching
for inconsistencies indicating effects from physics beyond the Standard
Model. Investigations of B hadrons are expected to break new ground in
measuring CP violation effects.

CP violation can occur in any theory in which there are complex
coefficients in the Lagrangian. In the Standard Model (SM), the origin of
the CP violation is the complex coupling of quarks to the Higgs field. After
the electroweak symmetry breaking, and after removing unphysical phases, one
irremovable complex phase remains in the three-generation quark mixing
matrix, the Cabibbo-Kobayashi-Maskawa (CKM) matrix. The matrix is defined
as:

\[ V=\left[ 
\begin{array}{lll}
V_{ud} & V_{us} & V_{ub} \\ 
V_{cd} & V_{cs} & V_{cb} \\ 
V_{td} & V_{ts} & V_{tb}
\end{array}
\right] \]

\[ =\left[ 
\begin{array}{lll}
1-\frac{1}{2}\lambda ^{2} & \lambda & A\lambda ^{3}(\rho -i\eta ) \\ 
-\lambda \left[ 1+A^{2}\lambda ^{4}(\rho +i\eta )\right] 
& 1-\frac{1}{2} \lambda ^{2} & A\lambda ^{2} \\ 
A\lambda ^{3}\left[ 1-(\rho +i\eta )(1-\frac{1}{2}\lambda ^{2})\right] & 
-A\lambda ^{2}\left[ (1-\frac{1}{2}\lambda ^{2})+\lambda ^{2}(\rho +i\eta
)\right] & 1
\end{array}
\right] \]

where the latter matrix is the Wolfenstein approximation to order $\lambda
^{5}$. $\lambda $ is the Cabibbo angle and is about 0.22, $A\simeq 1$ and 
$\rho \neq 0.$ The complex elements are $V_{ub}$ and $V_{td}$, if 
$\eta \neq 0.$

In general, there are three sources of CP violation: CP violation in decay
(`direct CP violation'), where the decay amplitude for a process and its'
complex conjugate are not the same; CP violation in mixing (`indirect CP
violation'), which occurs when two neutral mass eigenstates are not CP
eigenstates. CP violation can also originate from interference between
direct decays and decays via mixing, if neutral B and ${\rm \bar{B}}$ mesons
decay into the same final CP eigenstate. All these types of CP violation
have been observed in kaon decays, most recent case being the direct CP
violation, which the KTeV experiment could observe, thus confirming the
earlier results of NA31. Preliminary results from NA48 are in agreement with
the KTeV and NA31 measurements (for a review, see Ref. \cite{calvetti} in
these proceedings). Kaon decays have provided a wealth of CP violation
measurements, but they have not been able to probe the complex elements of
the CKM matrix, and thus the SM origin of the CP violation. For this
purpose, observing CP violation in B decays is mandatory.

The B mesons are expected to exhibit large CP-violation effects in some
(rare) decay modes, which probe the so-called Unitarity Triangle. Unitarity
of the CKM matrix defines triangle relations. The one which relates the
first and the third column of the CKM matrix turns out to form a triangle
in which all the sides are of the same order of magnitude:

\[ V_{ud}V_{ub}^{*}+V_{cd}V_{cb}^{*}+V_{td}V_{tb}^{*}=0. \]

\begin{figure}[htbp]
\mbox{\epsfig{figure=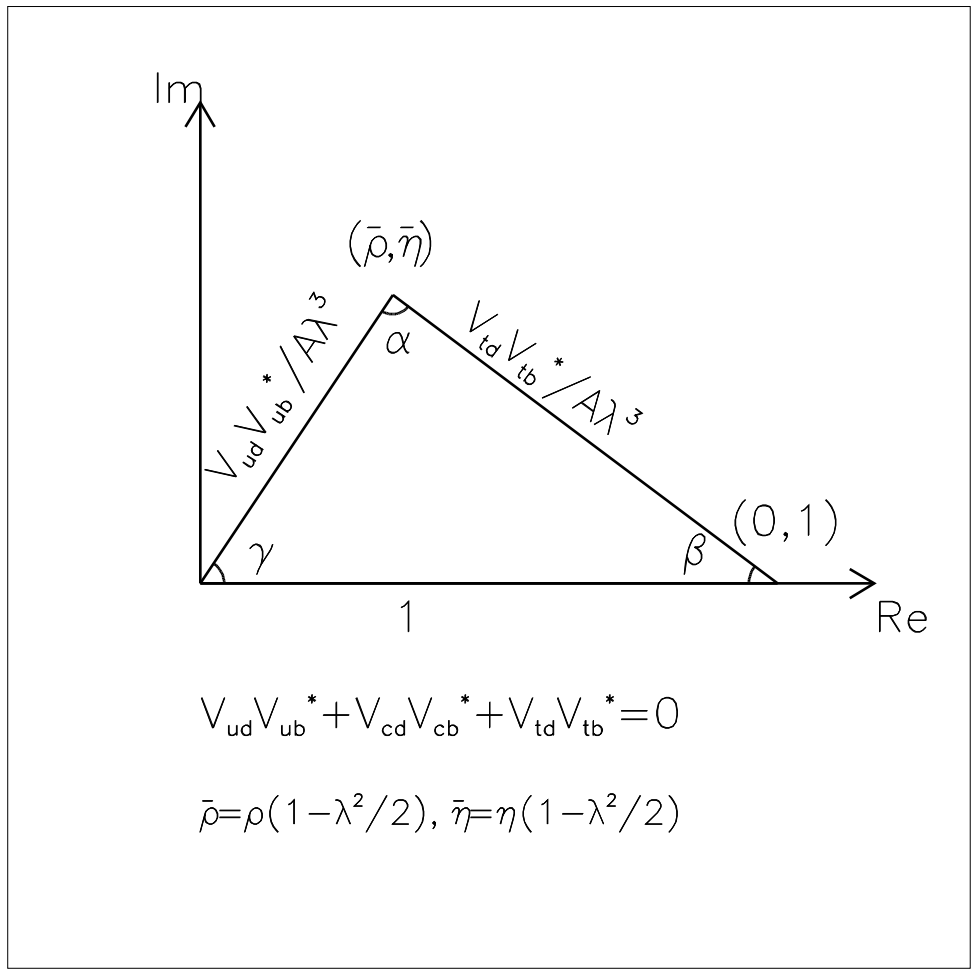,width=16cm}}
\caption{Unitarity triangle 
$V_{ud}V_{ub}^{*}+V_{cd}V_{cb}^{*}+V_{td}V_{tb}^{*}=0$. }
\label{tri}
\end{figure}

\begin{figure}[htbp]
\mbox{\epsfig{figure=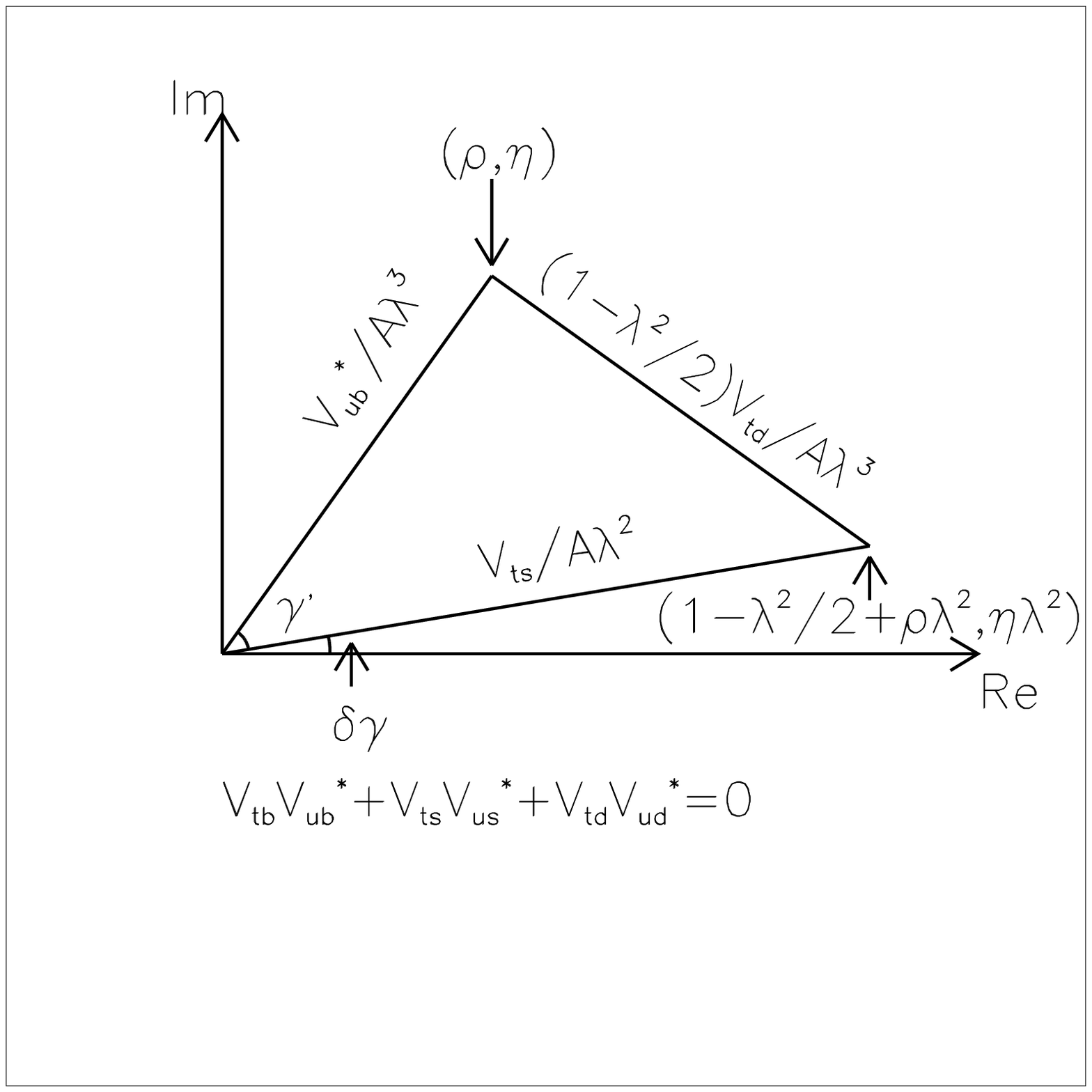,width=16cm}}
\caption{Unitarity triangle 
$V_{tb}V_{ub}^{*}+V_{ts}V_{us}^{*}+V_{td}V_{ud}^{*}=0$. }
\label{tri2}
\end{figure}

When all the sides are divided by $|V_{cd}V_{cb}^{*}|\simeq A\lambda ^{3}$
so that the base of the triangle is along the real axis and normalized to 1,
the apex of the triangle is located at $(\bar{\rho},\bar{\eta})$, where 
$\bar{\rho}=\rho (1-\lambda ^{2}/2)$ and 
$\bar{\eta}=\eta (1-\lambda ^{2}/2)$. 
One side of the triangle is related to $V_{ub}^{*}$, 
$V_{ud}V_{ub}^{*}/V_{cd}V_{cb}^{*}\simeq 0.4$, and the other side is related
to $V_{td},V_{td}V_{tb}^{*}/V_{cd}V_{cb}^{*}\sim 1$. The angles $\alpha $, 
$\beta $ and $\gamma $ are defined in Fig.~\ref{tri}. Another triangle can be
defined by combining the first and the third rows of the CKM matrix, and it
is shown in Fig.~\ref{tri2}. It probes the same two CKM matrix elements as
the `standard' Unitarity Triangle, and it's angles $\gamma ^{\prime }$ and 
$\delta \gamma $ can be measured.

When neutral B and ${\rm \bar{B}}$ mesons decay into the same final CP
eigenstate, the CP violation can in general have contributions both from
direct CP violation and from interference between direct and mixed decays.
In this case, the time-dependent CP asymmetry can be expressed as follows:

\[ a_{CP}(t)=\frac{\Gamma ({\rm B}_{{\rm q}}^{{\rm 0}}(t)\rightarrow 
{\rm f})-\Gamma ({\rm \bar{B}}_{{\rm q}}^{{\rm 0}}(t)\rightarrow 
{\rm f})}{\Gamma ({\rm B}_{{\rm q}}^{{\rm 0}}(t)\rightarrow 
{\rm f})+\Gamma ({\rm \bar{B}}_{{\rm q}}^{{\rm 0}}(t)\rightarrow 
{\rm f})} \]

\[ \qquad \ =A_{CP}^{dir}({\rm B}_{{\rm q}}^{{\rm 0}}\rightarrow 
{\rm f})\cos(\Delta m_{q}t)+A_{CP}^{int}({\rm B}_{{\rm q}}^{{\rm 0}}
\rightarrow {\rm f})\sin (\Delta m_{q}t), \]

where $A_{CP}^{dir}$ is the direct CP-violation amplitude and $A_{CP}^{int}$
is the mixing-induced CP-violation amplitude. Subscript q indicates a d- or
an s-quark, and $\Delta m_{q}$ is the mass difference of the 
B$_{{\rm q}}^{{\rm 0}}$ eigenstates. 
Here it is assumed that there is no width difference
of the B$_{{\rm q}}^{{\rm 0}}$ eigenstates, which is valid for 
B$_{{\rm d}}^{0}$ mesons. For B$_{{\rm s}}^{0}$, however, 
the width difference can be
sizeable, 15\% or so, and the formula has to be modified accordingly.

If the decay of a neutral B meson is dominated by a single CKM amplitude,
the mixing-induced CP-violation amplitude is $A_{CP}^{int}=\Lambda \sin
(\phi _{M}-\phi _{D})$. The mixing phase $\phi _{M}$ is $+2\beta $ in case
of B$_{{\rm d}}^{{\rm 0}}$ mesons and $-2\delta \gamma $ for 
B$_{{\rm s}}^{{\rm 0}}$ mesons, and $\Lambda $ is the CP eigenvalue 
of the final state.
The CP-violating weak decay phase $\phi _{D}$ is zero for dominant 
$\bar{b}\rightarrow \bar{c}cs(d)$ amplitudes, and 
$-2\gamma $ for dominant $\bar{b}\rightarrow \bar{u}ud(s)$ amplitudes.

\section{Measurement methods}

In the past few years, much theoretical progress has been made in
understanding the different sources of CP violation, and analysis methods
have been developed accordingly. Here, the measurement methods of angles 
$\beta $, $\alpha $ and $\gamma $ are described, with an emphasis on the
proposed new modes. These issues are described in more detail in the
theoretical contributions to this conference, see \cite{wyler}, 
\cite{Fleischer} and \cite{pirjol}. 
Analysis methods are also addressed briefly.

\subsection{Angle $\beta $}

The angle $\beta $ can be measured cleanly using the decays 
${\rm B_{d}^{0}\rightarrow J/\psi K_{S}^{0}}$. This decay is dominated by the
decay amplitude $\bar{b}\rightarrow \bar{c}cs$, since the penguins are
expected to be small and moreover, they have the same weak phase as the
tree-level decays. Therefore, the direct CP-violation contribution is very
small and can be neglected, $i.e.$ $A_{CP}^{dir}\simeq 0$, and the
mixing-induced CP-violation amplitude is $A_{CP}^{int}=-\sin 2\beta $.

\subsection{Angle $\alpha$}

The angle $\alpha $ can be measured with the decay 
${\rm B_{d}^{0} \rightarrow \pi ^{+}\pi ^{-}}$. 
This decay has, however, contributions from
both penguin and tree decay graphs, and recent CLEO measurements of 
${\rm B\rightarrow \pi \pi ,K\pi }$ 
indicate that the penguins play a significant
role. If there were no penguin contributions present, the mixing-induced
CP-violation amplitude would be 
$A_{CP}^{int}=\sin (2\beta +2\gamma )=-\sin 2\alpha $, 
assuming the SM triangle relation $\alpha +\beta +\gamma =\pi $.
Including the penguin contributions the amplitudes can be written in the
form:

\[ A_{CP}^{dir}=2\frac{A_{P}}{A_{T}}\sin \delta \sin \alpha , \]

\[ A_{CP}^{int}=-\sin 2\alpha -2\frac{A_{P}}{A_{T}}\cos \delta \cos 2\alpha
\sin \alpha , \]

where $A_{P}/A_{T}$ is the ratio of the penguin and tree amplitudes, and 
$\delta $ is the strong phase difference between the amplitudes. If the 
ratio 
$A_{P}/A_{T}$ can be predicted accurately with the help of additional
branching ratio measurements, $\alpha $ and $\delta $ can be extracted from
the measured asymmetry.

There is an alternative proposal of measuring the angle $\alpha $ using
decays ${\rm B_{d}^{0}\rightarrow \rho \pi }$ {\rm \cite{rhopi}}. This
option has been studied recently for example by the BaBar collaboration 
\cite{babarbook}. Another way of attacking the problem is to use isospin
relations between the decays ${\rm B_{d}^{0}\rightarrow \pi ^{+}\pi ^{-}}$, 
${\rm B_{d}^{0}\rightarrow \pi ^{0}\pi ^{0}}$, 
${\rm B^{+}\rightarrow \pi^{+}\pi ^{0}}$ and 
the charge conjugated modes \cite{isospin}. In this way
the angle $\alpha $ could be measured unambiguously. The decay mode 
${\rm B_{d}^{0}\rightarrow \pi ^{0}\pi ^{0}}$ is, 
however, predicted to have a
branching ratio of $10^{-6}$ or less, which makes the method experimentally
difficult.

\subsection{Angle $\gamma $}

The angle $\gamma $ is the most difficult to measure, since there is no both
experimentally feasible and theoretically clean decay of a B to a CP\
eigenstate, which could be used (the equivalent channel to 
${\rm B_{d}^{0}\rightarrow J/\psi K_{S}^{0}}$ and 
${\rm B_{d}^{0}\rightarrow \pi^{+}\pi ^{-}}$ would be 
${\rm B_{s}^{0}\rightarrow \rho K_{S}^{0}}$). There
has been, however, a considerable amount of theoretical activity recently to
find new methods of probing the angle $\gamma $.

A method of measuring the angle $\gamma $ is to measure the decay time
distributions of ${\rm B_{s}^{0}\rightarrow D_{s}^{\pm }K}^{{\rm \mp }}$ and
the charge conjugated modes \cite{dsk}. Here CP violation is due to
interference of direct and mixed decays. Tagging is obviously needed to
distinguish ${\rm B_{s}^{0}}$ and ${\rm \bar{B}_{s}^{0}}$. Fitting the two
decay asymmetries $vs.$ decay time gives $\gamma -2\delta \gamma $ and 
$\delta $, where $\delta $ is the strong phase, and $\delta \gamma $ is the
phase of $V_{ts}.$

Another method of measuring the angle $\gamma $ is to consider the decays 
${\rm B_{d}^{0}\rightarrow D_{CP}^{0}K^{*}}$, 
${\rm B_{d}^{0}\rightarrow \bar{D^{0}}K^{*}}$, 
${\rm B_{d}^{0}\rightarrow D^{0}K^{*}}$ and their charge
conjugated modes \cite{dunietz}. The amplitudes of these decays are related
by two triangles in the complex plane. The triangles differ in the length of
one side only, and an angle between the two triangles is $2\gamma $. In
addition to the weak phase, there is a dependence on a strong phase
difference $\delta $, but it can be extracted from the measurements. Similar
relations exist for the charged B decays: ${\rm B^{+}\rightarrow
D_{CP}^{0}K^{+}}$, ${\rm B^{+}\rightarrow \bar{D^{0}}K^{+}}$, 
${\rm B_{d}^{0}\rightarrow D^{0}K^{+}}$ and charge conjugated modes 
\cite{gronauwyler}. These decays are self-tagging, so no external tag is
required. On the other hand, these decays are purely hadronic, requiring a
hadronic first level trigger. Furthermore, some of the decay modes have
small branching ratios, which make the measurements difficult, in particular
for BaBar and BELLE.

The decay ${\rm B_{d}^{0}\rightarrow D^{(*)\pm }\pi ^{\mp }}$has been
proposed already a while ago for angle $\gamma $ measurement \cite{dstarpi}.
Even if these decays are not decays to CP eigenstates, both ${\rm B_{d}^{0}}
$ and ${\rm \bar{B}_{d}^{0}}$ can decay to the same final state, leading to
interference between mixing and decay, which measures $\sin (2\beta +\gamma
) $, and a strong phase difference $\delta .$ When the angle $\beta $ will
have been measured accurately in the ${\rm B_{d}^{0}\rightarrow J/\psi
K_{S}^{0}}$ decays, the angle $\gamma $ can be extracted. These decays were
first payed little attention due to problems with statistics -- even if the
branching fractions are not that small, of the order of $10^{-3}$, one of
the decay paths is doubly Cabibbo-suppressed and therefore the CP-violating
effects are tiny. Furthermore, the reconstruction efficiency is small for
fully reconstructed final states. The channel was, however, recently
re-considered by BaBar \cite{babarbook}, for which the channel would provide
an access to the angle $\gamma $ since ${\rm B_{d}^{0}}$ decays are
involved. They realized that the statistics can be improved considerably by
using inclusive reconstruction. The channel has also been studied by LHCb
collaboration with promising results, indicating an accuracy of four degrees
for the angle $\gamma $ after five years of data-taking \cite{schneider}.

A lot of theoretical effort has been put recently in learning how to extract
angle $\gamma $, or set limits to $\gamma ,$ using various combinations of 
${\rm K\pi }$, ${\rm \pi \pi }$, and ${\rm KK}$ final states. Here these
modes are addressed only briefly, more details can be found elsewhere in
these proceedings \cite{wyler}, \cite{Fleischer}, \cite{pirjol}. Three
different combinations of these decays have been proposed to set limits to
the angle $\gamma $: ${\rm B^{\pm }\rightarrow \pi ^{\pm }K}$ and 
${\rm B_{d}^{0}\rightarrow \pi ^{\mp }K^{\pm }}$ \cite{fleishermannel}, 
${\rm B^{\pm }\rightarrow \pi ^{\pm }K}$ and 
${\rm B^{\pm }\rightarrow \pi^{0}K^{\pm }}$ \cite{gronaurosnerlondon}, 
\cite{burasfleischer}, and ${\rm B_{d}^{0}\rightarrow \pi ^{0}K}$ 
and ${\rm B_{d}^{0}\rightarrow \pi ^{\mp}K^{\pm }}$ \cite{burasfleischer}. 
The strategies are based on
flavour-symmetry arguments, although the theoretical understanding of the
hadronic final state interaction effects is poor at the moment.

The strategies of simultaneously determining $2\beta $ and $\gamma $ using
decays ${\rm B_{d}^{0}\rightarrow \pi ^{+}\pi ^{-}}$ and 
${\rm B_{s}^{0}\rightarrow K^{+}K^{-}}$ was also 
discussed in this conference \cite{Fleischer}; 
using this combination of decays is theoretically cleaner,
since the theoretical accuracy is only limited by U-spin breaking effects
(interchanging d and s-quarks). Similar arguments were presented for decays 
${\rm B_{s(d)}^{0}\rightarrow J/\psi K_{S}^{0}}$ and 
${\rm B_{d(s)}^{0}\rightarrow D}_{{\rm d(s)}}^{+}{\rm D_{d(s)}^{-}.}$

\subsection{Measurements of ${\rm B_{s}^{0}}$-mesons}

The ${\rm B_{s}^{0}}$-meson mixing parameter $\Delta m_{s}$ has not been
measured yet, due to the rapid oscillations of ${\rm B_{s}^{0}}$-mesons. The 
${\rm B_{s}^{0}}$-meson mixing gives an important constraint to the
Unitarity Triangle. One side of the triangle is proportional to $V_{td}$,
which is related to the ${\rm B_{d}^{0}}$-meson mixing parameter $\Delta
m_{d}.$ Despite of the fact that there is a rather precise measurement of 
$\Delta m_{d}$, several hadronic uncertainties limit the precision of the 
$V_{td}$ measurement. In the ratio $\Delta m_{s}/\Delta m_{d}$, on the other
hand, many common factors such as QCD corrections and dependence on the top
quark mass cancel, and a more accurate estimation of $V_{td}$ is possible.
The ratio can be written as

\[ \frac{\Delta m_{s}}{\Delta m_{d}}=\frac{m({\rm B}_{{\rm s}})}
{m({\rm B}_{{\rm d}})}\frac{B({\rm B}_{{\rm s}})f^{2}
({\rm B}_{{\rm s}})}{B({\rm B}_{{\rm d}})f^{2}({\rm B}_{{\rm d}})}
\frac{|V_{tb}^{*}V_{ts}|}{|V_{tb}^{*}V_{td}|}, \]

where $m({\rm B}_{q})$ are the B-meson masses, $B({\rm B}_{{\rm q}})$ are
the bag parameters, and $f({\rm B}_{{\rm q}})$ the B-meson form factors. In
order to measure the ${\rm B_{s}^{0}}$-mixing, the decay time has to be
measured accurately, and the flavour of the B has to be tagged both at the
decay time (tagged by the observed decay itself) and at production. To tag
the flavour of the B at the production time, an external tag has to be used
-- these tagging methods are discussed in more detail in the following
section. The asymmetry between mixed and non-mixed decays is:

\[ A(t)=D\frac{\cos (\Delta m_{s}t)}{\cosh (\Delta \Gamma _{s}t/2)}, \]

where the dilution factor $D$ comes from finite proper time resolution,
mistags, and background. $\Delta \Gamma _{s}=\Gamma _{H}-\Gamma _{L}$ is the
width difference between the ${\rm B_{s}^{0}}$ mass eigenstates ${\rm B_{H}}$
and ${\rm B_{L}.}$ Oscillations are most often searched for using methods
based on fitting the amplitude of the oscillatory term, keeping the $\Delta
m_{s}$ fixed, and repeating the fit with different values of $\Delta m_{s}$ 
\cite{amplitude}. This method has the advantage that results from different
experiments can be easily combined.

The decay channel ${\rm B_{s}^{0}\rightarrow J/\psi \phi }$ is very useful
for extracting various, as yet unmeasured or poorly measured parameters of
the ${\rm B_{s}^{0}}$ mesons. Furthermore, the decay channel is
experimentally an easy one since it can be triggered and reconstructed
cleanly. The ${\rm B_{s}^{0}}$ decay proper time and the angular
distributions of the secondary particles can be used for extracting $\Delta
\Gamma _{s}$, $\Gamma _{s}=(\Gamma _{H}+\Gamma _{L})/2$, and CP amplitudes 
$A_{||}$ and $A_{T}$ describing the decays to CP-even and CP-odd
configurations. The angular analysis is sensitive to the strong phase
differences between the amplitudes as well. All these parameters can
measured by using untagged samples.

The weak phase of the decay ${\rm B_{s}^{0}\rightarrow J/\psi \phi }$ is
proportional to $2\delta \gamma $, and it is very small in the SM.
Furthermore, the weak phase measurement requires an external tag.
Nevertheless, this decay mode can be used for constraining the angle 
$\gamma $ together with the various other decay modes. 
In particular, larger than
expected CP violation would indicate that processes beyond the SM are
involved.

\subsection{Analysis methods}

Experimental analysis methods have been improved considerably over the
years. In particular, many new or complementary tagging methods have been
established; in addition to the traditional opposite-side lepton tagging and
kaon tagging, hadronic tagging methods such as same-side pion
tagging, same- or opposite side jet-charge tagging, and vertex charge
tagging have shown their power in increasing the statistics significantly.
The CDF experiment, for example, has demonstrated that the poorer purity of
the hadronic tags is compensated by the bigger efficiency, and the tagging
quality factor $\epsilon D^{2}$ ($\epsilon $ = efficiency, $D$ = dilution)
is roughly the same for the opposite side lepton tags (2.2\%) as for the
same-side hadron tags (2.1\%) and the jet-charge tags (2.2\%) 
\cite{maksim}. 
The combined use of these methods has allowed the experiments at LEP, SLC
and Tevatron to push the limits of their B-physics programmes by improving
considerably the tagging efficiency. Measurements of $\Delta m_{d}$, limits
on the ${\rm B_{s}^{0}}$-mixing, and the asymmetry measurement of tagged 
${\rm B_{d}^{0}\rightarrow J/\psi K_{S}^{0}}$ decays in particular, are
examples of results unforeseen some years ago.

In the early days of planning CP-violation measurements using B-decays,
expected performance was evaluated by simply estimating event rates. More
thorough analysis methods are needed to extract all the available
information. For example, time-dependent measurement of CP-asymmetries is
almost mandatory, not only to prove the expected time-dependence of the
asymmetry, but also to help in distinguishing the signal from backgrounds
with different time-dependence. For mixing measurements, amplitude fits have
provided a useful platform at LEP, SLC and Tevatron to extract and combine
experimental limits. Angular analyses will have to be utilized for channels
in which the B-meson decays into spin-carrying non-stable particles, such as
in the case of ${\rm B_{s}^{0}\rightarrow J/\psi \phi }$ decay.

\section{Present results}

\subsection{Direct measurement of angle $\beta$}

\begin{figure}[htbp]
\mbox{\epsfig{figure=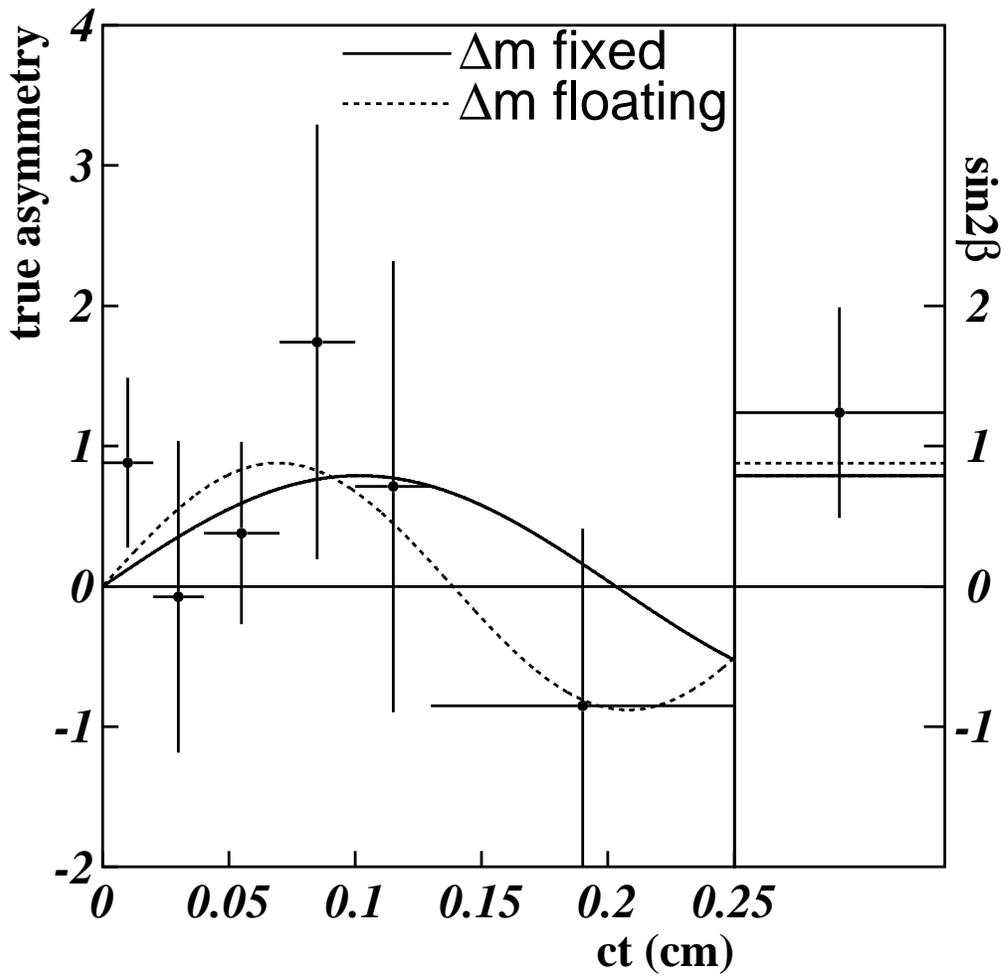,width=14cm}}
\caption{CDF measurement of asymmetry in ${\rm B_{d}^{0}\rightarrow J/\psi
K_{S}^{0}}$ decays \protect\cite{maksim}. }
\label{cdf_asym}
\end{figure}

The CDF limit on angle $\beta $ was reported in this 
conference \cite{maksim}. The ${\rm B_{d}^{0}\rightarrow J/\psi K_{S}^{0}}$ 
sample consisted of about 200 events in which both muons from ${\rm J/\psi }$ 
are within the SVX acceptance, and about 200 events with one or no muons 
reconstructed in the SVX. Data were collected at the Tevatron Run I, and 
consisted of an integrated luminosity of 110 pb$^{-1}$. The events with both 
muons in the SVX had a precise lifetime information, and a time-dependent 
analysis could be performed. The additional sample of 200 events had 
imprecise lifetime information. A time-averaged analysis could be performed, 
however, improving the precision of the combined measurement. Soft leptons, 
jet-charge, and same-side hadrons were used for tagging, each providing a 
similar tagging quality of about 2.2\%. The combined tagging quality factor 
was (6.3$\pm $1.7)\%, accounting for correlations and double tags. The 
tagging efficiency and dilution factors were measured using an independent sample of 
${\rm B^{+}\rightarrow J/\psi K^{+}}$ decays, and a 
${\rm B\rightarrow \ell D}^{(*)}{\rm X}$ sample suitably scaled to account 
for momentum differences. The result of the full unbinned maximum likelihood 
fit was: $\sin 2\beta =0.79_{-0.44}^{+0.41}.$ The measured asymmetry is 
shown in Fig.~\ref{cdf_asym}. Interpreting the measurement as a limit to 
$\sin 2\beta $, CDF quote $0<\sin 2\beta <1$ at 93\% CL using the 
Feldman-Cousins frequentist approach, and at 95\% CL using the Bayesian 
approach.

The OPAL experiment at LEP performed the first exploratory search for CP
asymmetry in ${\rm B_{d}^{0}\rightarrow J/\psi K_{S}^{0}}$ decays, using a
sample of 24 reconstructed events, out of which 10 events were estimated to
be background \cite{opal}. The measurement yielded 
$\sin 2\beta=3.2_{-2.0}^{+1.8}\pm 0.5.$

\subsection{Precision measurements}

The CKM matrix element ratios $|V_{ub}/V_{cb}|$ and $|V_{td}/V_{cb}|$ are
related to the lengths of the sides of the Unitarity Triangle. In this
conference, updated results were presented by CLEO on $|V_{cb}|$ and 
$|V_{ub}/V_{cb}|$ \cite{cleoresults}, yielding

\[ |V_{cb}|=(38.5\pm 2.1\pm 2.2\pm 1.2)\cdot 10^{-3}, \]
\[ |V_{ub}|=(3.25\pm 0.14_{-0.29}^{+0.21}\pm 0.55)\cdot 10^{-3}, \]

where the last error is the theoretical uncertainty. The LEP experiments
have made progress in measuring these matrix elements in the LEP1 data, with
different systematical uncertainties from CLEO. New results on $|V_{cb}|$
were presented in this conference from analyses using exclusively
reconstructed final states, see Ref. \cite{lepresults}; the errors of
individual experiments are approaching the ones from CLEO. Averaging over
the LEP results, including both the inclusive $|V_{cb}|$ determinations
from lifetime and branching ratio measurements and the exclusive
measurements, yields \cite{vcbtampere}:

\[ |V_{cb}|=(40.2\pm 1.9)\cdot 10^{-3}. \]

The LEP average for $|V_{ub}|$ reported in this conference was:

\[ |V_{ub}|=(4.03_{-0.46}^{+0.39}\pm 0.56)\cdot 10^{-3}, \]

where the last error is the theoretical uncertainty. The matrix element 
$V_{td}$ is at the moment constrained by the $\Delta m_{d}$ measurements and
the $\Delta m_{s}$ limits. The ${\rm B_{d}^{0}}$ meson oscillation frequency 
$\Delta m_{d}$ has been measured accurately by the LEP experiments, SLD and
CDF, while the $\Upsilon $(4S) experiments provide time-integrated mixing
measurements. The LEP average value reported in this conference yields a
value of $\Delta m_{d}=(0.468\pm 0.019)$ ps$^{-1}$ \cite{lepresults}, while
the world average including LEP, SLD, CDF and $\Upsilon $(4S) experiments
is $\Delta m_{d}=(0.473\pm 0.016)$ ps$^{-1}$ \cite{bosc}. The combined LEP
95\% CL limit for the ${\rm B_{s}^{0}}$ meson oscillation frequency 
$\Delta m_{s}$ is 9.6 ps$^{-1}$, while the sensitivity is 12.6 ps$^{-1}$. 
The world
average limit by the LEP experiments, SLD, and CDF is 14.3 ps$^{-1}$, and
the sensitivity is at 14.7 ps$^{-1}$ \cite{bosc}.

Lifetimes of ${\rm B^{+}}$, ${\rm B_{d}^{0}}$, ${\rm B_{s}^{0}}$, 
${\rm B_{c}}$ and B-baryons were reported by CDF and the LEP experiments. 
The combined
results were \cite{maksim}, \cite{lepresults}:

\[ \tau ({\rm B^{+}})=(1.66\pm 0.03) \: {\rm ps}, \] 
\[ \tau ({\rm B_{d}^{0}})=(1.55\pm 0.03)\: {\rm ps}, \] 
\[ \tau ({\rm B_{s}^{0}})=(1.47\pm 0.06) \: {\rm ps}, \] 
\[ \tau ({\rm B_{c})}=(0.46\pm 0.17) \: {\rm ps} \: {\rm (CDF)}, \]
\[ \tau (\Lambda {\rm _{b})}=(1.23\pm 0.08) \: {\rm ps},  \] 
\[ \tau (\Xi _{{\rm b}}^{0}{\rm )}=(1.39_{-0.28}^{+0.34})\: {\rm ps} \: {\rm (ALEPH,} 
\: {\rm DELPHI)}. \] 

Very recently, new branching ratio measurements were presented by CLEO 
\cite{jaffe}. The measured branching ratio for the decay 
${\rm B_{d}^{0}\rightarrow \pi ^{+}\pi ^{-}}$, 
BR(${\rm B_{d}^{0}\rightarrow \pi^{+}\pi ^{-}}$) 
$=(4.7_{-1.5}^{+1.8}\pm 1.3)\cdot 10^{-6}$, indicates
difficulties for the angle $\alpha $ measurements, in particular at the 
${\rm e}^{{\rm +}}{\rm e}^{{\rm -}}$ B-factories. The much larger branching
ratio for the decay ${\rm B_{d}^{0}\rightarrow K^{-}\pi ^{+}}$, 
BR(${\rm B_{d}^{0}\rightarrow K^{-}\pi ^{+}}$) 
$=(1.88_{-0.26}^{+0.28}\pm 0.06)\cdot 10^{-5}$, 
suggests large penguin contributions. CLEO is also probing the
way of constraining the angle $\gamma $ by comparing the rates for charged
and neutral B decays to ${\rm K\pi }$ using the Neubert-Rosner analysis.
CLEO results \cite{cleoresults} favour $\gamma >90^{\circ }$, a result
which would be in contradiction with the Unitarity Triangle apex region
allowed by the $\Delta m_{s}$ limit.

CDF presented complementary results on the ${\rm B_{s}^{0}}$ frontier 
\cite{maksim}. Using a sample of about 600 inclusively reconstructed 
${\rm B_{s}^{0}\rightarrow \ell D}_{{\rm s}}{\rm X}$ decays, a two-component
lifetime fit of the form $\exp (-\Gamma _{H}t)+\exp (-\Gamma _{L}t)$ was
used to measure the width difference $\Delta \Gamma _{s}$, yielding a result 
$\Delta \Gamma _{s}/\Gamma _{s}=0.34_{-0.34}^{+0.31}.$ Interpreted as a
limit, the measurement yielded $\Delta \Gamma _{s}/\Gamma _{s}<0.83$ at 95\%
CL. In the SM, the width difference $\Delta \Gamma _{s}$ is related to 
$\Delta m_{s}$ by a hadronic scale factor. The CDF limit on upper limit on 
$\Delta \Gamma _{s}$ gives thus an upper limit of 96 ps$^{-1}$ to 
$\Delta m_{s}.$

CDF presented also polarisation measurements. The measurement of the
polarisation in ${\rm B_{d}^{0}\rightarrow J/\psi K}^{{\rm *0}}$ used about
200 events, similar to CLEO. The polarisation measurement of 
${\rm B_{s}^{0}\rightarrow J/\psi \phi }$ used about 
40 events -- it is the first
polarisation measurement of its kind, albeit limited by the statistics. With
a larger statistics, $\Delta \Gamma _{s}$ can be measured together with the
CP-even and CP-odd amplitudes $A_{||}$ and $A_{T}$, along with the strong
phases, as discussed in Section 2.4.

The precision data on $|V_{cb}|$, $|V_{ub}/V_{cb}|$ and $\Delta m_{d}$, 
the $\Delta m_{s}$ limit, and $\epsilon _{K}$ from kaon experiments, combined
with estimates on hadronic factors from mass measurements and lattice
calculations, can be used to constrain the Unitarity Triangle. Various such
fits have been performed, see for example Refs. \cite{stocchi}, \cite{mele}
and \cite{ali}. Theoretical errors for the $V_{ub}$ are already limiting the
bound from $|V_{ub}/V_{cb}|$. The direct measurement of angle $\beta $ is
completely consistent with the region allowed by the indirect measurements,
but the present accuracy of the $\sin 2\beta $ measurement does not improve
the overall fit. In Ref. \cite{stocchi}, the following ranges for the angles 
$\alpha $, $\beta $ and $\gamma $ are quoted:

\[ \sin 2\beta =0.725_{-0.060}^{+0.050}, \]
\[ \sin 2\alpha =-0.26_{-0.28}^{+0.29}, \]
\[\gamma =(59.5_{-7.5}^{+8.5})^{\circ }. \]

\begin{figure}[tbp]
\mbox{\epsfig{figure=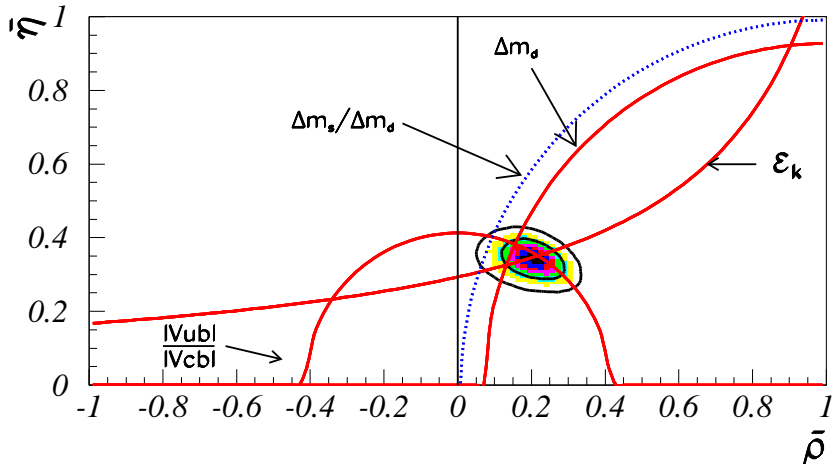,width=14cm}}
\caption{Unitarity Triangle constraints from present precision data (from
Ref. \protect\cite{stocchi}). }
\label{utfig}
\end{figure}

The region limited in the complex space by the combined fit is shown in 
Fig.~\ref{utfig}.

\section{Experiments embarking on data taking}

The timing of the conference was quite exciting, since there were several
B-experiments getting ready for data-taking. This conference was the first
one to see collision events from BaBar and BELLE. HERA-B experiment is
expected to be fully commissioned by end of 1999, and the upgraded CDF and
D0 experiments will re-start at the high-luminosity Tevatron around mid-2000.

\subsection{HERA-B}

\begin{figure}[tbp]
\mbox{\epsfig{figure=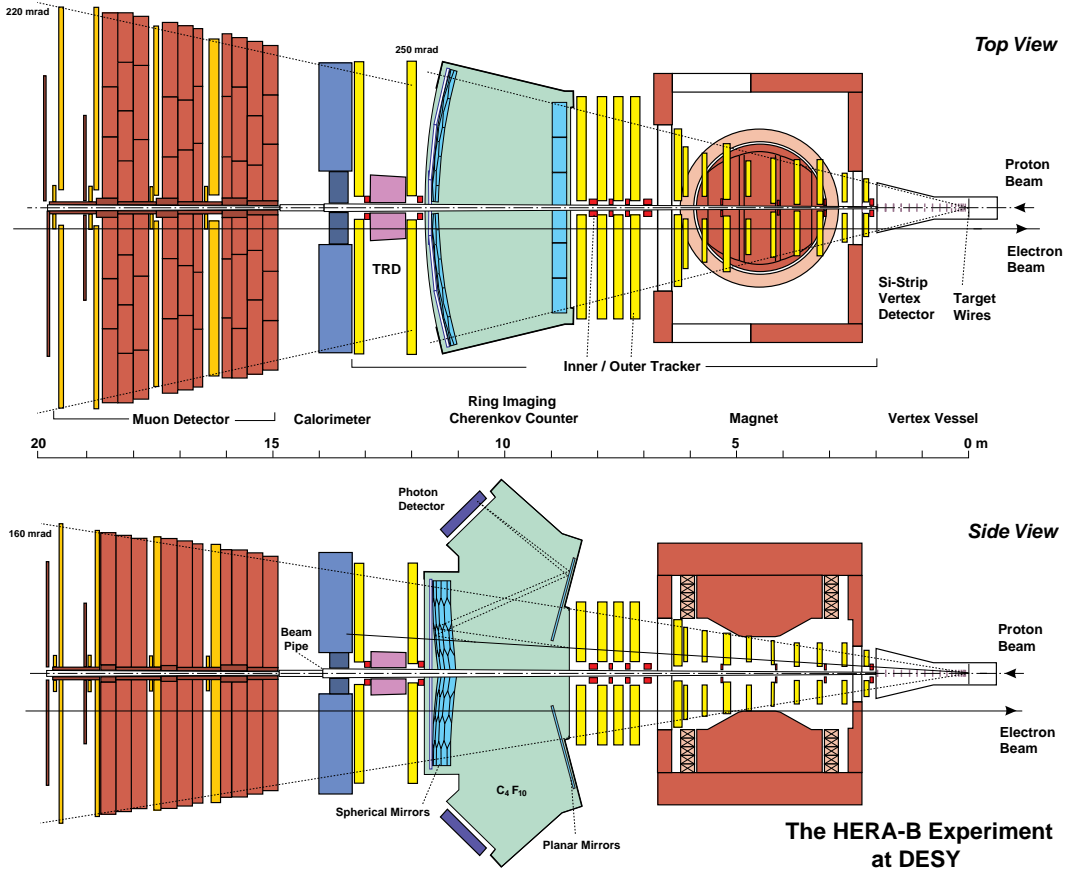,width=14cm}}
\caption{The HERA-B detector layout.}
\label{herabfig}
\end{figure}

HERA-B experiment at DESY is a fixed-target B-experiment, operating at the
920 GeV proton ring HERA \cite{padilla}. The detector layout is shown in
Fig. \ref{herabfig}. The experiment is challenging due to the small ratio of 
${\rm b\bar{b}}$-to-total cross-section, about 10$^{-6}$, resulting in
difficult trigger and target requirements. At the center-of-mass energy of
about 40 GeV, the ${\rm b\bar{b}}$ cross-section is about 12 nb, albeit with
large uncertainties. In order to obtain a competitive number of signal
events in the decay mode ${\rm B_{d}^{0}\rightarrow J/\psi K_{S}^{0}}$ (1500
reconstructed events in a year), four interactions per bunch crossing are
needed, resulting in an interaction rate of 40 MHz. This corresponds to an
integrated luminosity of about 30 fb$^{-1}$ in a year. The requirements to
the trigger, and the high radiation dose that some of the HERA-B
subdetectors will have to stand, are comparable to the conditions at LHC.

To maximise the interaction rate while minimising the interference to the
ep-operation at HERA, the target was designed to consist of two sets of four
wires placed at the beam halo, located at a distance of four to ten standard
deviations from the beam \cite{ehret}. The target has been in continuous
operation since 1997, and the main requirements have been achieved.

The HERA-B trigger consists of three levels \cite{xella}. The first level
trigger searches for two lepton candidates with an invariant mass compatible
with a J/$\psi $, or two high-$p_{{\rm T}}$ hadrons compatible with
originating from a B-meson. The track seeds originate from the muon pad
chambers, the electromagnetic calorimeter or from the high-$p_{{\rm T}}$
chambers. A fast hardware tracking is then performed, in which regions of
interest are scanned in the upstream detector layers using a Kalman
filter-type method. The track candidates are refined in the second level,
while the third level trigger uses data outside regions of interest as well.
The final output level (level 4) reconstructs and classifies the events
without any further rate reduction.

Many of the sub-detectors are completed and entering the full commissioning
phase: the vertex detector (VDS) \cite{wagner}, the Ring Imaging Cherenkov
Counter (RICH) \cite{pyrlik}, electromagnetic calorimeter (ECAL) 
\cite{zoccoli}, 
the TDR and the muon detector \cite{titov}. The data-acquisition
system has been running stable for several months, and the commissioning of
the first level trigger (FLT) is starting. Some HERA-B subdetectors have
been suffering from technological problems to fulfil the requirements of
operating in a high-radiation environment. The anode aging and Malter effect
problems with the outer tracker, which consists of honeycomb drift-chambers,
seem to have been solved by cathode Au-coating, proper gas choice and
careful validation of materials as well as manufacturing procedures. After
applying these measures, the detectors have been proven to survive in a
hadronic environment an equivalent radiation dose of two integrated HERA-B
years \cite{capeans}. The mass production is progressing in a rapid pace,
and all the chambers are planned to be installed by November 1999. The inner
tracker is using MicroStrip Gas Chambers (MSGCs). Original problems with
sparking were solved by adding Gas Electron Multipliers (GEMs) into the
drift space. Even though the chambers still show a gain variation as a
function of the accumulated charge, the detectors have been considered as
adequate for HERA-B \cite{herab-msgc}. The mass production is proceeding and
the whole inner tracker is expected to be completed by the end of 1999.

\subsection{${\rm e}^{{\rm +}}{\rm e}^{{\rm -}}$ B-factory experiments}

The BaBar experiment at SLAC and the BELLE experiment at KEK are asymmetric 
${\rm e}^{{\rm +}}{\rm e}^{{\rm -}}$ collider experiments, 
collecting ${\rm B\bar{B}}$ pairs produced in $\Upsilon $(4S) decays. 
The production
cross-section of b${\rm \bar{b}}$ pairs is about 1 nb. Both experiments took
their first data just prior to this conference, and exciting first event
views were shown.

In BaBar, the 9 GeV electron and 3.1 GeV positron beams, circulating in the
double PEPII ring, are brought to collide with a zero-angle crossing, which
requires strong magnets within the interaction region to deflect the beams
between bunches. The nominal luminosity is $3\cdot 10^{33}$~cm$^{-2}$s$^{-1}$
-- the nominal one year data corresponds thus to 30 fb$^{-1}$. The BaBar
detector has to fulfil special requirements on its' tracking parts due to
the asymmetric collisions: since the boost is along the beam axis, the
difference of the decay times of the two B-mesons is measured by the
difference in the $z$ component of the vertex separation. Furthermore, the 
$p_{{\rm T}}$-range of the final state particles is quite low, between 60 MeV
and 4 GeV. The tracking system consists of a Silicon Vertex Tracker (SVT)
and a drift chamber (DCH), immersed in a 1.5 T solenoidal field. The charged
hadrons are identified in the Detector for Internally Reflected Cherenkov
light (DIRC), which is a novel technology using quartz bars both for
producing the Cherenkov light and as light guides \cite{hocker}. The
electromagnetic calorimeter is a CsI calorimeter, and outside the EMC the
Instrumented Flux Return (IFR) is used for muon identification and for
enhancing ${\rm K_{L}^{0}}$ detection. The collaboration is using software
written entirely in C++.

\begin{figure}[htbp]
\mbox{\epsfig{figure=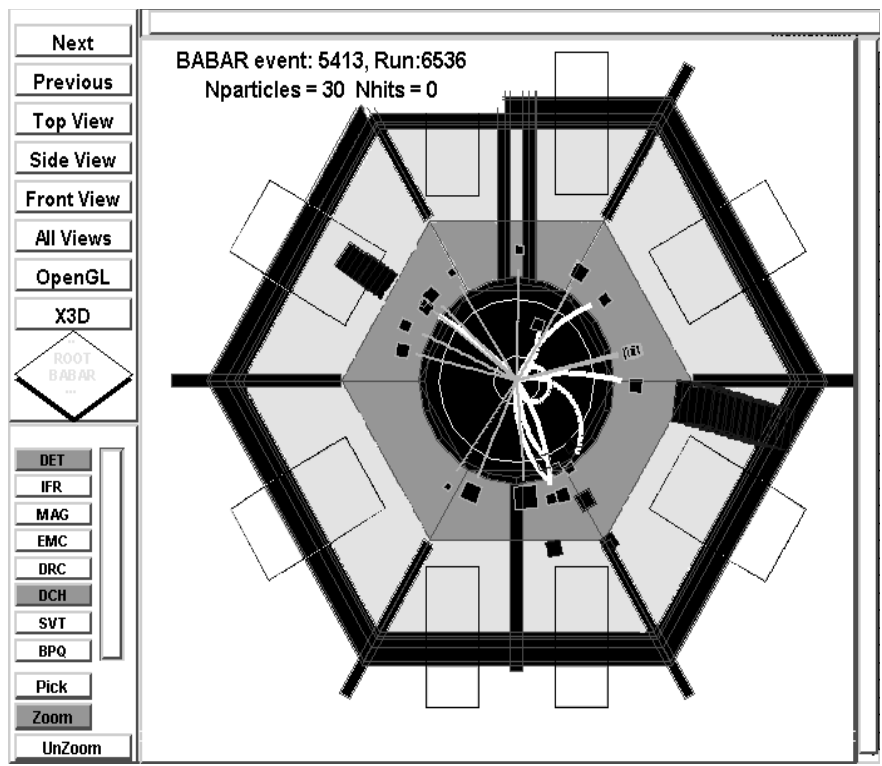,width=14cm,height=11cm}}
\caption{A hadronic event reconstructed in BaBar.}
\label{babarfig}
\end{figure}

BaBar started taking colliding beam data May 26$^{{\rm th}}$ 1999 
\cite{McMahon}. 
By June 17$^{{\rm th}}$ 1999, time of the conference, 10 pb$^{-1}$
were collected on tape and first hadronic events were shown. An example of a
reconstructed hadronic event is shown in Fig. \ref{babarfig}. Mass peaks for 
${\rm K_{S}^{0}\rightarrow \pi ^{+}\pi ^{-}}$ with a width of 6 MeV (loose
vertex pointing requirements, no particle ID) and ${\rm \pi ^{0}\rightarrow
\gamma \gamma }$ were reconstructed and shown. In addition, $60\cdot 10^{6}$
cosmics had been collected prior to the colliding beam data taking. All
subsystems were operational and complete apart from a fraction of the DIRC
bars. Since the conference, the exclusive B reconstruction was under way and
the first ${\rm B_{d}^{0}\rightarrow J/\psi K_{S}^{0}}$ candidates were
identified.

The BELLE experiment is operating at the KEKB ring, which produces
collisions between 8 GeV electron and 3.5 GeV positron beams. The ring
design includes several novel features, such as a finite-angle beam
crossing, which has the advantage that there is no need for strong magnets
within the interaction region, making the design of the tracker detectors
easier. The nominal design luminosity is $1\cdot 10^{34}$~cm$^{-2}$s$^{-1}$.
If bunch-bunch instabilities appear, a crab-crossing scheme, in which the
bunches are tilted so that they collide head-on despite of the finite
crossing angle, will be a possible solution. The tracking system consists of
a Silicon Vertex Detector (SVD) and a central drift chamber (CDC) in a 1.5
T solenoidal field. The charged hadrons are identified by using a
time-of-flight (TOF) counter and an aerogel Cherenkov counter (ACC). The
electromagnetic calorimeter consists of CsI crystals and the instrumented
iron yoke serves as a muon and ${\rm K_{L}^{0}}$ detector.

\begin{figure}[htbp]
\mbox{\epsfig{figure=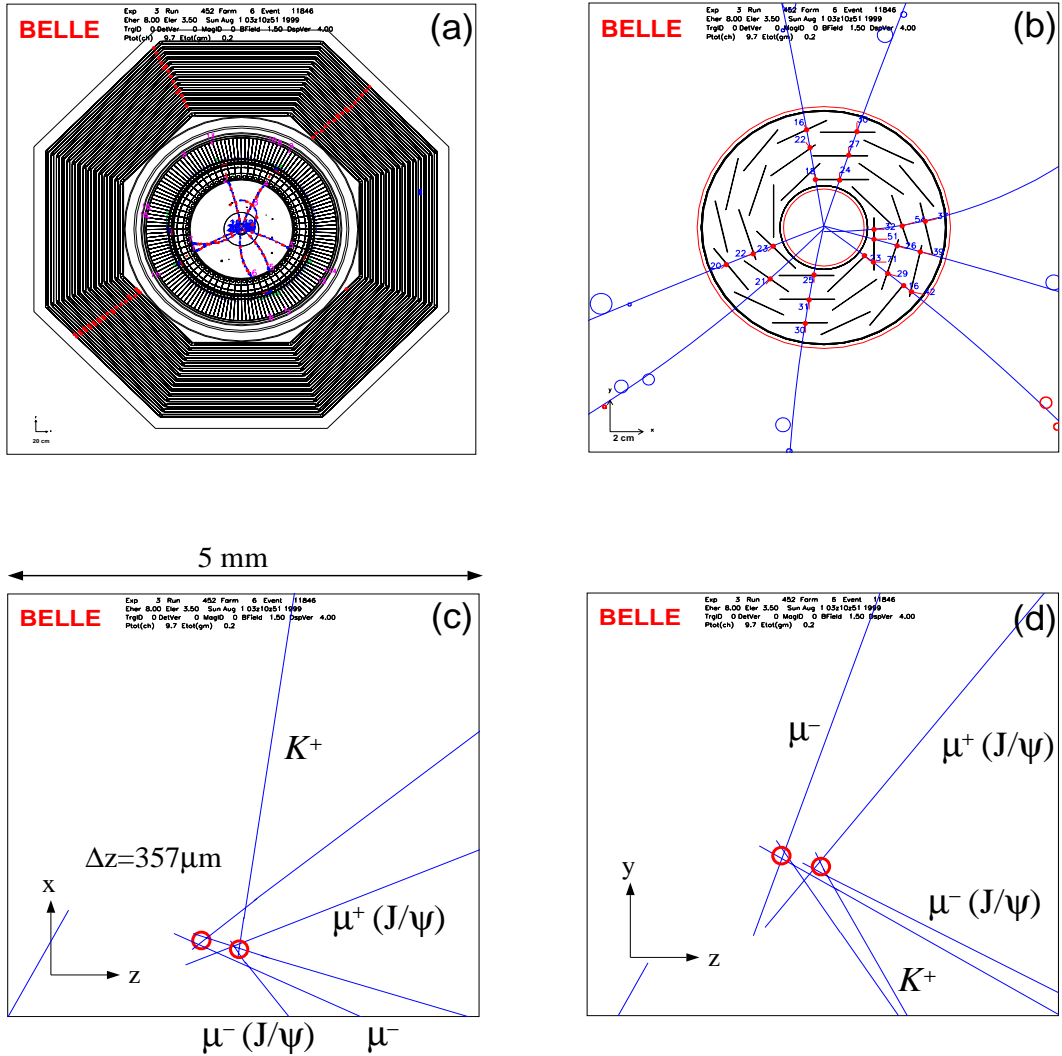,width=14cm}}
\caption{A hadronic event reconstructed in BELLE.}
\label{bellefig}
\end{figure}

The first collision events were observed in June 1$^{{\rm st}}$ 1999 
\cite{kek}. 
By June 10$^{{\rm th}}$ 1999, an integrated luminosity of about 
530 nb$^{-1}$ was collected. 
In this conference, reconstructed 
${\rm J/\psi\rightarrow e^{+}e^{-}}$ and 
${\rm \mu \mu }$ events were shown, as well as
mass peaks for ${\rm K_{S}^{0}\rightarrow \pi ^{+}\pi ^{-}}$ and for 
${\rm \pi ^{0}\rightarrow \gamma \gamma .}$ 
The ${\rm K_{S}^{0}}$ mass resolution
was 2.5 MeV. The resolution for the ${\rm \pi ^{0}}$ was about 5.6 MeV, with 
$E_{\gamma}>$~20 MeV. An example of a reconstructed hadronic
event is shown in Fig.~\ref{bellefig}. After the conference, till 
August 4$^{{\rm th}}$ 1999, 
a total of 25 pb$^{-1}$ of data was recorded by BELLE, with
the accelerator reaching a peak luminosity of 
$2.9\cdot 10^{32}$ cm$^{-2}$s$^{-1}$. 
Most of the data were taken on the $\Upsilon $(4S) peak.

The CLEOIII detector is an upgrade of the CLEO detector at CESR to achieve
luminosities in excess of 10$^{33}$ cm$^{-2}$s$^{-1}$ \cite{berkelman}.
CESR is a symmetric ${\rm e}^{{\rm +}}{\rm e}^{{\rm -}}$ collider operating
at the $\Upsilon $(4S) peak. The main features of the upgrade concerns the
particle identification and tracking \cite{viehhauser}. Charged hadron
identification will be enhanced with a RICH detector. The tracking system
had to be rebuilt due to the reduced space available -- the new tracker will
consists of a new drift chamber, and a silicon vertex detector, whose main
function is in providing precision tracking. The CLEOIII experiment will
start operating during 1999, and it is aiming at a wide range of B-physics
accessible with rate measurements.

\subsection{Upgraded CDF and D0}

The CDF and D0 experiments at the Tevatron ${\rm p\bar{p}}$ collider are
being upgraded for the high-luminosity run (RunII). The RunII is scheduled
to begin in the summer of 2000, with an increased luminosity resulting from
many improvements in the collider, the main one being the replacement of the
old Main Ring with the new Main Injector. The crossing rate will be 132~ns
for 121~bunches, corresponding to a typical luminosity of 
1.62$\cdot 10^{32}$~cm$^{-2}$~s$^{-1}$ and a 
peak luminosity of 2$\cdot 10^{32}$~cm$^{-2}$~s$^{-1}$. 
At the peak luminosity, the average number of interactions per
crossing is two. The design goal is to collect an integrated luminosity of
2~fb$^{-1}$ in about two years of running. An increase of the collision
energy from 1.8~TeV to 2.0~TeV is planned as well, driven by the 
40\% enhancement in the ${\rm t\bar{t}}$ yield. 
At Tevatron, the ${\rm b\bar{b}}$
production cross-section is about 100~$\mu $b. The ratio of the 
${\rm b\bar{b}}$-to-total cross-section is about 10$^{-3}$, 
requiring efficient trigger strategies.

The CDF detector will have many new features to improve the resolution and
radiation tolerance, and to cope with the shorter crossing rate 
\cite{papadimitri}. 
There will be a new seven-layer radiation hard silicon vertex
and tracking system, extending between radii 2.4~cm to 28~cm (1.6~cm with
the innermost Layer00, beyond the baseline detector). The central outer
tracker, COT, will be new as well, since the available radial space will be
less, and the maximum drift time has to be shorter than the crossing time.
The calorimeters beyond $|\eta |>1$ will be replaced with a scintillating
tile calorimeter, and a new muon system will be built to cover the region 
$1.0<|\eta |<1.5$. The trigger and data acquisition system will be fully
pipelined. Fast tracking will be available at level-1, allowing for an
all-hadronic B-decay trigger. In the level-2, it will be possible to trigger
on tracks with a high impact-parameter. Beyond the baseline design, a TOF
detector has been recommended to enhance particle identification at very low
momenta (0.3 GeV $<p_{{\rm T}}<$1.6 GeV).

The D0 detector is undergoing a major upgrade, driven by both new physics
goals and changes in the collider \cite{Lucotte}. The whole tracking system
is new, consisting of a Central Fiber Tracker (CFT) and a Silicon Microstrip
Tracker (SMT) in a 2~T solenoidal field. The electron identification is
enhanced by adding preshower detectors in front of the electromagnetic
calorimeter, and the muon system has been upgraded significantly. To cope
with the shorter bunch spacing while covering the physics goals, the
front-end electronics, trigger and data-acquisition systems have been
largely re-designed. The level-1 B-decay triggers are based on soft leptons
-- hadron or vertex triggers are not considered. In addition to the single
and di-muon triggers, ${\rm e^{+}e^{-}}$ pairs can be triggered on by
matching either a track element in the outermost CFT layers ($p_{{\rm T}}$
threshold 1.5 GeV) and a calorimeter cluster ($E_{{\rm T}}$ threshold 1.5
GeV), or a preshower cluster (threshold 2-5 MIPs) and a calorimeter cluster.

\section{Experiments being designed and constructed}

By the year 2005, various experiments will have explored the Unitarity
Triangle. It is likely that the angle $\beta $ will have been measured with
a fair precision, $\sigma (\sin 2\beta )$=0.05 or so. Sides of the Unitarity
Triangle will have been measured: $V_{ub}$, possibly still limited by
hadronic uncertainties, and $V_{td}$ from mixing measurements of both 
${\rm B_{d}}$ and ${\rm B_{s}}$. 
It is still quite possible that for the angle 
$\alpha $, there is only a low statistics measurement, with theoretical
uncertainties, and for the angle $\gamma $, there is no accurate measurement
or even no direct measurement at all. Considering the allowed region for the
apex of the Unitarity Triangle, the angle $\beta $ is actually not
constraining it significantly, but it is the angle $\gamma $ which is
significantly reducing the size of the allowed region and thus placing a
stringent test to the SM. Therefore, the next generation of experiments at
LHC and at Tevatron are needed for full investigation of CP violation in
order to over-constrain the CKM matrix via high statistics measurements,
complemented by theoretically clean channels.

\subsection{Dedicated B-physics experiments LHCb and BTeV}

\begin{figure}[htbp]
\mbox{\epsfig{figure=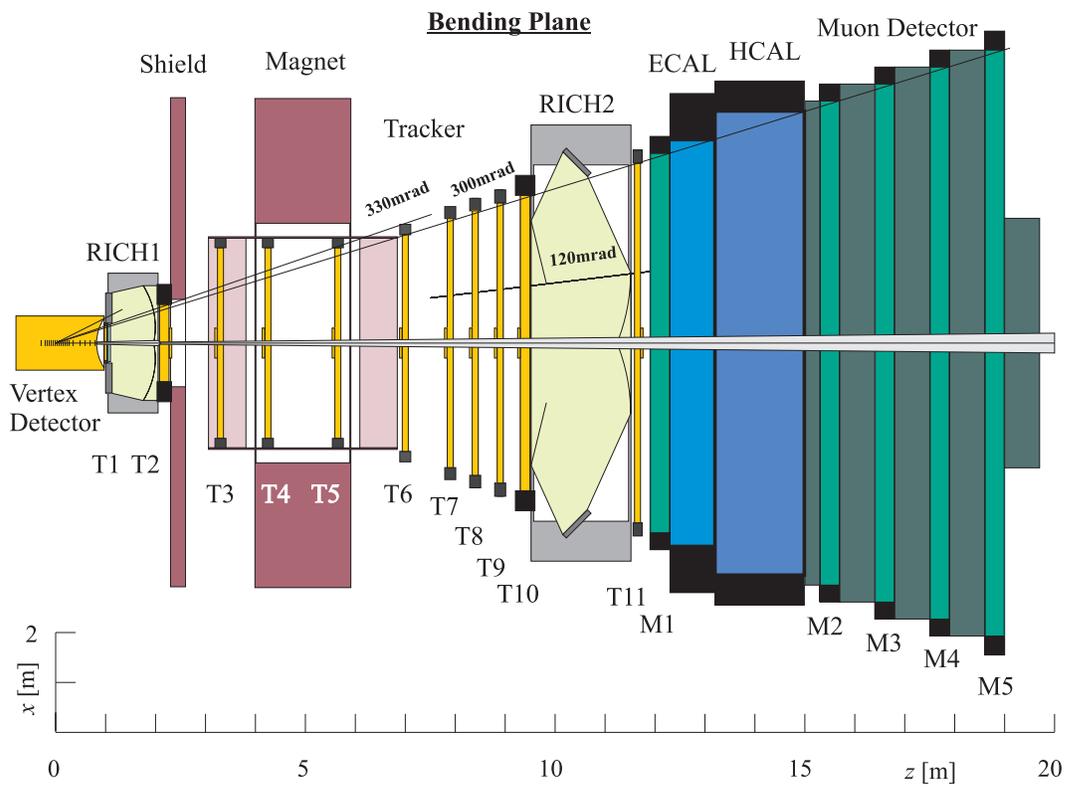,width=14cm}}
\caption{The LHCb detector layout.}
\label{lhcbfig}
\end{figure}

LHCb is a forward single-arm spectrometer at LHC with efficient trigger,
good particle identification and good momentum resolution \cite{schneider}, 
\cite{lhcb}. The LHCb detector layout is shown in Fig. \ref{lhcbfig}. The
main components of the detector are a vertex detector inside the beam
vacuum, a tracking system with a dipole magnet, two RICH counters to
identify charged hadrons over a sufficiently large momentum range,
calorimeters, and a muon system. The angular coverage is from 15 mrad to 300
mrad, corresponding to about three units in pseudo-rapidity. The single-arm
geometry exploits the fact that the ${\rm b\bar{b}}$ system is boosted along
the beam, and thus the B and the ${\rm \bar{B}}$ mesons are highly
correlated in pseudo-rapidity.

At LHC, the total ${\rm b\bar{b}}$ cross-section is about 500~$\mu $b, with
an uncertainty of a factor of two or more. The fraction of 
${\rm b\bar{b}}$-to-total cross-section is of order $10^{-2}$. 
At LHCb, the LHC beam is
`detuned' to produce a constant luminosity of 
$2\cdot 10^{32}$~cm$^{-2}$s$^{-1}$. 
This luminosity is chosen to optimize the fraction of single pp
interactions per bunch crossing to minimize the radiation damage, and to
minimize the detector occupancy to make pattern recognition easier. At the
luminosity of $2\cdot 10^{32}$~cm$^{-2}$s$^{-1}$, the fraction of bunches
with a single pp interaction is about 30\%.

At LHCb, the level-0 trigger selects first single pp interactions by
applying a pile-up veto, obtained from two dedicated Si-disks. Single
interactions are then passed on to be processed by $p_{{\rm T}}$ or 
$E_{{\rm T}}$ single track triggers, 
provided by muon chambers for muons (20\% of
events), ECAL for electrons/photons (10\%), or ECAL and HCAL for hadrons
(60\%). The rate is reduced from 40 MHz to 1 MHz.

The level-1 trigger is based on topological identification of secondary
vertices. Track elements are found in the Si-disks, the primary vertex is
reconstructed, and secondary vertices are formed by using large impact
parameter tracks. The rate is expected to be reduced from 1 MHz to 40 kHz.
The higher level triggers (levels two and three) will consist of software
algorithms running on farms of commercial computers. The higher level
triggers will have to reject uninteresting ${\rm b\bar{b}}$ events in
addition to non-b events. The target rate for data recording is 200 Hz.

\begin{figure}[htbp]
\mbox{\epsfig{figure=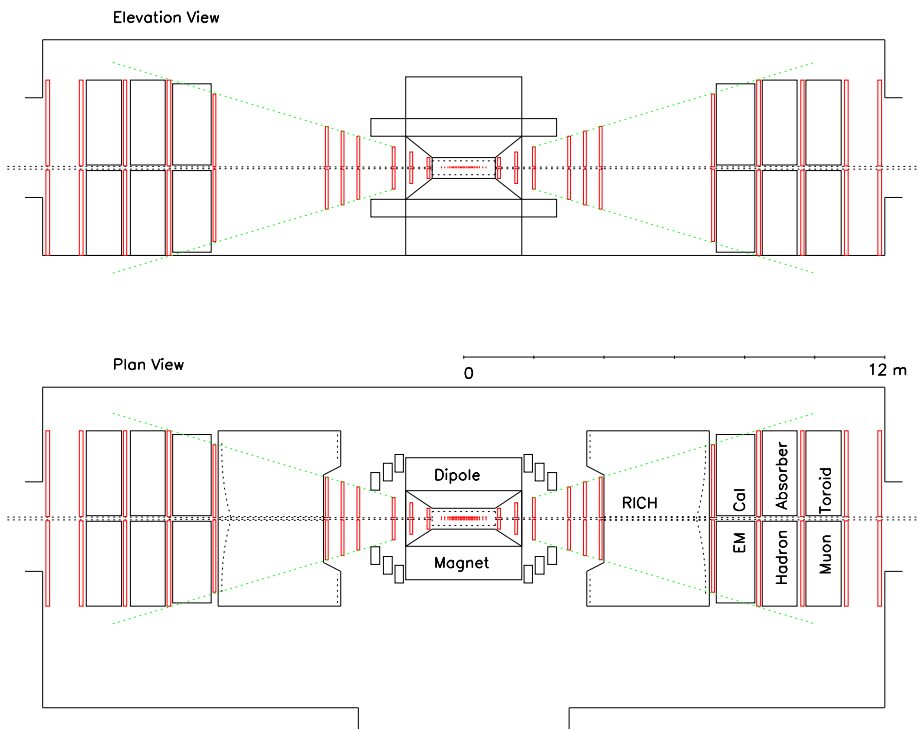,width=14cm}}
\caption{The BTeV detector layout.}
\label{btevfig}
\end{figure}

BTeV is a forward two-arm spectrometer proposed for Tevatron \cite{btev}, 
\cite{gardner}. The BTeV detector layout is shown in Fig. \ref{btevfig}. The
detector is planned to have a good particle identification, and in
particular, it is relying on a secondary vertex level-1 trigger by using
track vectors from pixel triplets. The main components of the detector are a
pixel vertex detector located inside a dipole magnet on the interaction
region, forward trackers, RICH counters, high-resolution PbWO$_{4}$
electromagnetic calorimeters, hadron absorbers and muon system consisting of
toroid spectrometers. The angular coverage is up to 300 mrad on both arms.
The total ${\rm b\bar{b}}$ cross-section is a factor of five less than at
LHC; however, there are many compensating factors working in favour of BTeV.
The two-arm solutions brings an obvious factor of two. Another factor of two
comes from running with two interactions per crossing, which is possible at
Tevatron due to the long luminous region which makes it easier to separate
different primary vertices, and the longer bunch spacing (132 ns).
Furthermore, the smaller boost of B hadrons increases the track angles with respect
to the beam pipe, thus reducing the radiation damage and making the pattern
recognition easier. The smaller boost also enables BTeV to manage with a
single RICH detector, whereas LHCb has to use two separate RICHes to cover
the whole dynamical range. The overall length of the detector can be made
shorter as well.

A vertex trigger at the level-1 is naturally very desirable to trigger
efficiently on purely hadronic final states. To achieve this, the
tracking/vertex detector design has to be tied closely to the trigger
design. The baseline is to use pixel triplets in a dipole field, thus
forming track vectors. Primary vertex can be formed from these track
vectors, and large impact parameter tracks can be searched for. Preliminary
studies indicate that requiring at least two tracks detached by more the
four standard deviations from the primary vertex, only 1\% of the beam
crossings are triggered, while an efficiency from 40\% to 70\% can be
achieved for the most commonly studied hadronic decays 
(${\rm B^{0}\rightarrow h^{+}h^{-}}$, 
${\rm \ B_{s}^{0}\rightarrow D_{s}K}$, 
${\rm B^{-}\rightarrow D^{0}K^{-}}$, 
${\rm B^{-}\rightarrow K_{S}^{0}\pi ^{-}}$, 
${\rm B_{d}^{0}\rightarrow J/\psi K_{S}^{0}}$, 
${\rm B_{d}^{0}\rightarrow J/\psi K^{*}}$, 
${\rm B^{0}\rightarrow K^{*}\gamma }$ ) \cite{gardner}.

\subsection{General purpose experiments ATLAS and CMS at LHC}

The general purpose experiments ATLAS and CMS at LHC are high-$p_{{\rm T}}$
central collider experiments, designed primarily for heavy particle
searches, but with a significant capacity for precision physics, in
particular in the first years of LHC operation with a nominal `low'
luminosity of $1\cdot 10^{33}$~cm$^{-2}$s$^{-1}$, giving an integrated
luminosity of 10 fb$^{-1}$ per year \cite{atlastdr}, \cite{jones}, 
\cite{starodumov}. 
The limiting factor for the maximum luminosity for B physics
is the level-1 trigger output rate, rather than the offline reconstruction
capability. Some channels such as rare decays of the type ${\rm B\rightarrow
\mu \mu (X)}$ are particularly suitable for ATLAS and CMS since they are
self-triggering, and the high-luminosity data of LHC can be used as well.

The angular coverage of the tracking systems of these detectors is about
five units in pseudo-rapidity. Since the ${\rm b\bar{b}}$ production
cross-section is roughly flat in pseudo-rapidity, the production rate is as
high in the central as in the forward region. On the other hand, the small
boost of central b quarks, and the consequent lack of angular correlation
between the b and ${\rm \bar{b}}$, make triggering of B-hadron final states
difficult. The lack of single hadron triggers with sufficiently low 
$p_{{\rm T}}$ at the lowest trigger level, 
and the poor charged hadron identification
are the most obvious handicaps of the general purpose experiments with
respect to dedicated B-physics experiments.

Both ATLAS and CMS have powerful muon systems which are capable of
triggering on single low-$p_{{\rm T}}$ muons at the lowest trigger level --
the threshold for a single muon trigger is 6 GeV for ATLAS and 7 GeV for
CMS. The single muon trigger makes the backbone of the B-physics triggers.
CMS can use a lower threshold of 2-4 GeV for the level-1 muon if other
signals are available in addition to the muon. After applying the trigger
level-1 acceptance, the number of triggered ${\rm b\bar{b}}$ events is
typically few times 10$^{10}$ per year. At the level-2 and at higher trigger
levels, the triggers are designed to select specific B-decay channels.

ATLAS and CMS have pixel detectors as the detector element closest to the
interaction point, thus providing a good vertex resolution. The large
tracking systems enable efficient charged particle reconstruction. In
addition, ATLAS has the capability for identifying low-$p_{{\rm T}}$
electrons ($p_{{\rm T}}>0.5$ GeV) by using the transition radiation in the
TRT, while both experiments can use their electromagnetic calorimeters for
identifying soft electrons at somewhat higher $p_{{\rm T}}.$ The use of the
ATLAS TRT for $\pi $/K separation has been investigated recently 
\cite{barberis}. 
The results are very encouraging and would enhance greatly the
analysis of the two-hadron decays of B-mesons, however, the results are
still to be confirmed at beam tests.

\section{Detector R\&D, prototypes and running experience}

\subsection{Targets and beams}

In this conference, first running experience was reported on the PEPII and
KEKB asymmetric ${\rm \ e}^{{\rm +}}{\rm e}^{{\rm ^{-}}}$ beam operation 
\cite{kozanecki}, \cite{iijima}. At KEKB, synchrotron light background from
the whole ring was found to be a significant issue for the BELLE operation,
requiring further work. At PEPII, the main source of BaBar background was
found to be lost particles, coming from local bremsstrahlung, and local or
distant Coulomb scattering.

HERA-B experiment has already several years experience on the wire target
operation \cite{ehret}. Two sets of four wires are placed at the beam halo,
located at a distance of four to ten standard deviations from the beam. The
distance is not constant, but as the beam intensity drops, the wires are
brought continuously closer to the beam core to keep a constant interaction
rate. Even though the proton beam lifetime is reduced due to the fact that
the wire target operation is consuming between 1 and 2 mA/h of the proton
current, the luminosity and the background conditions to the other HERA\
experiments are not affected. A reliable target steering and stable
operation have been achieved.

\subsection{Multilevel triggering}

Apart from the ${\rm e}^{{\rm +}}{\rm e}^{{\rm -}}$ B-experiments, the
trigger is a vital issue since the fraction of B-events varies from about
1\% (LHC) down to 10$^{-6}$ (HERA-B). Moreover, since the branching
fractions of interesting final states are typically small, the experiments
have to have interaction rates ${\cal O}$(10 MHz) or more. All the
experiments use a multilevel triggering scheme. Some experiments use a
pre-trigger (level-0) either to discard empty crossings (HERA-B), or to make
a pile-up veto (LHCb). The lowest level physics trigger (level-1) is
searching for high-$p_{{\rm T}}$ leptons or hadrons. This is achieved in
most experiments by using fast trigger detectors stand-alone. Only HERA-B
will use tracking in level-1 to confirm the high-$p_{{\rm T}}$ hadron
signals. BTeV is the only experiment which is planning to use an event
topology trigger at level-1, searching for displaced vertices. At the higher
trigger levels, data is kept in pipelines for online event processing. Data
storage is expected to operate at a 100~Hz rate, apart from BTeV which is
planning to compress the data online, and to write out only data summary
information.

\subsection{Tracking and vertexing}

Microvertex detectors, located as close to the primary vertex as practically
possible, are crucial for precise secondary vertex measurements. 
In ${\rm e^{+}e^{-}}$ experiments at the $\Upsilon $(4S), 
the vertex detectors
operate also as stand-alone tracking devices for very low-momentum
particles. In forward spectrometer geometries, retractable detector disks
are located inside the beam vacuum vessel, while in central experiments the
closest detector layer can be placed just outside the beam pipe. Technologies
employed are either Si-strip detectors or Si-pixels. Radiation hardness is
an issue even for the vertex detectors in ${\rm e^{+}e^{-}}$ experiments --
for example the BaBar vertex detector is foreseen to tolerate a 2.4 kGy dose
per year, thus guaranteeing a 10-year operation \cite{McMahon}. For
comparison, at HERA-B, in the detector area closest to the beam, the
maximum dose per year is 100 kGy, and it is foreseen that the silicon
detectors will be exchanged once per year \cite{wagner}.

Tracking systems are facing unforeseen operational requirements with the
high particle rates. Progress has been made in solving detector physics and
chemistry problems at high particle rates. HERA-B experience with the
honeycomb drift chambers has given valuable lessons on the careful choice of
materials, glues and gases \cite{capeans}. The problems at HERA-B with a new
device, MSGC, are not that surprising since these detectors have never been
used before in high-energy physics experiments. While the HERA-B experiment
has equipped their MSGCs with GEMs \cite{herab-msgc}, thus achieving
satisfactory performance, the CMS experiment is studying using advanced
passivation on their MSGCs \cite{latronico}. In this method, a strip of
dielectric material is deposited over all the cathode edges to prevent
electron extraction by the high electric field at the strip edges. Using
this method, together with high surface resistivity and narrow anodes, the
CMS prototype detectors have been proven to operate in an LHC-like
radiation environment.

The possibility for $dE/dx$ measurements has proven to be a valuable asset,
if no other particle identification method is available and if the momentum
spectrum of particles is suitable. It has been used in ${\rm e^{+}e^{-}}$
experiments (CLEO, LEP experiments) as well as in hadronic experiments
(CDF).

\subsection{Calorimetry and muons}

The electromagnetic calorimeters are used for triggering electrons and
photons, and for reconstructing electrons, photons, and $\pi ^{0}$ and $\eta 
$ mesons. Different experiments have very different dynamic ranges and
radiation hardness requirements, leading to different choices for
calorimeters. The ${\rm e^{+}e^{-}}$ experiments (CLEO, BaBar, BELLE)
operate at low energies, and the choice has been CsI crystals, which are
sensitive down to 10-20 MeV photons. Hadron experiments require a large
dynamic range and high radiation hardness. The choice has been either
high-resolution crystal calorimeters (PbWO$_{4}$) in BTeV and CMS, or
standard sampling calorimeters with more modest energy resolution
(lead-scintillators in LHCb, lead-LAr in ATLAS). The hadron calorimeters
play a minor role in B-physics. Nevertheless, they are used for muon
identification and as a muon filter, as well as for hadron triggers and for
enhancing ${\rm K_{L}^{0}}$ identification in BaBar and BELLE.

Muons represent the cleanest way of tagging B-meson final states, and all
experiments are designed to trigger and reconstruct muons with a high
efficiency. Some experiments (D0, BTeV, ATLAS) have a powerful stand-alone
muon momentum measurement, provided by toroids in the muon system.

\subsection{Charged hadron identification}

Charged hadron identification is a feature which often puts contradictory
requirements on the detector, compromising the operation of other detector
elements. Therefore dedicated devices such as Cherenkov counters are only
incorporated in experiments which consider physics with identified hadrons
as their first priority. This is obviously the case with dedicated B-physics
experiments, which all are relying on Cherenkov counters of some sort. In
addition, Cherenkov counters are used extensively in heavy-ion and
astrophysics experiments. General purpose high-energy physics experiments
have to compromise between the requirements on the particle identification
on one hand, and on the minimisation of material in front of the
calorimeters and maximisation of the tracking volume, on the other hand.
Therefore, most general purpose experiments have chosen to have either a 
$dE/dx$ capability in the tracker (CDF, ATLAS), or no identification at all
-- among the recent general purpose collider experiments, only DELPHI and
SLD have had Cherenkov counters.

Cherenkov counters are the most practical solution for identifying charged
hadrons over a large momentum range, typically between 1 and 150 GeV. 
Relativistic particles emit Cherenkov
radiation in a medium. The angle of emission is proportional to the velocity
of particle as $\theta =\arccos (1/n\beta )$, where $n$ is the index of
refraction. When the angle is measured using photosensitive devices, and the
momentum is measured independently in another detector, the mass of the
particle can be defined. The choice of radiator medium depends on the
desired momentum range for identification, and on the available space. The
larger the refractive index, the larger the emission angle, and the radiator
can thus be made thinner. On the other hand, large refractive index is
correlated with a high density, which is harmful for the operation of other
subdetectors. Characteristics of radiators in Cherenkov detectors in
B-physics experiments are summarised in Table~\ref{radiator}.

\begin{table}[htb]
\centering
\begin{tabular}{|l|c|c|c|c|c|}
\hline
Radiator & CF$_4$ & C$_4$F$_{10}$ & Aerogel & LiF & fused silica \\ 
material &  &  &  &  & (quartz) \\ \hline
$n$ & 1.0005 & 1.0014 & 1.01-1.10 & 1.50 & 1.474 \\ 
&  &  & (ref. 1.03) & ($\lambda=150$ nm) &  \\ \hline
$\theta^{{\rm {max}}}_{{\rm {c}}}$ [mrad] & 32 & 53 & 242 & 841 & 825 \\ 
$p_{{\rm {thr}}}^\pi$ [GeV] & 4.4 & 2.6 & 0.6 & 0.12 & 0.13 \\ 
$p_{{\rm {thr}}}^{{\rm {K}}}$ [GeV] & 15.6 & 9.3 & 2.0 & 0.44 & 0.46 \\ 
State & gas & gas & solid & solid & solid \\ \hline
Experiment & LHCb & HERA-B & BELLE & CLEOIII & BaBar \\ 
&  & BTeV & LHCb &  &  \\ 
&  & LHCb &  &  &  \\ \hline
\end{tabular}
\caption{Characteristics of radiators in RICH detectors in B-physics
experiments: refractive index $n$, maximum emission angle 
$\theta^{{\rm {max}}}_{{\rm {c}}}=\arccos(1/n)$, 
and threshold momenta for pions and kaons 
($p_{{\rm {thr}}}>m/\protect\sqrt{(n^2-1)}$). 
The state is given for room temperature.}
\label{radiator}
\end{table}

The light collection must be efficient to obtain a sufficient amount of
photoelectrons per particle. The detector walls have to have a high
transparency for the radiated photon spectrum (walls are typically made of
quartz, CaF$_{2}$, or borosilicate), while the detectors must have a high
quantum efficiency \cite{seguinot}, \cite{ekelof}. Detectors can be either
multiwire chambers with gas doped with TMAE or TEA as the photosensitive
component, multiwire chambers with reflective CsI photocathode, standard
PMTs, or hybrid photodiodes (HPDs), which combine a photocathode to a small
Si-strip or Si-pad drift chamber \cite{weilhammer}.

The specific ionisation ($dE/dx$) can be used at a more limited momentum range than
Cherenkov radiation for charged hadron identification. The
minimum of the Bethe-Bloch curve for the specific ionisation is around 
$\beta \gamma $=3.0-3.5, and the relativistic rise region of 
$\beta \gamma $
which can be used for particle identification is roughly from 10 to 400,
leading to $\pi $/K separation for momenta between 1 and 50 GeV 
\cite{dolgoschein}. Moreover, the required detector length is larger than 
for Cherenkov counters to achieve the same separation.
Particle identification by $dE/dx$ is thus a feasible option
for B-physics experiments which have a significant fraction of the B-hadron
decay products in this momentum range. This is the case for the 
${\rm e^{+}e^{-}}$ experiments at the $\Upsilon $(4S), 
and also the central
collider experiments, while at forward experiments the momentum spectrum is
harder.

The particle identification technologies used or foreseen for B-physics
experiments are summarized in Table~\ref{hid}.

\begin{table}[htb]
\centering
\begin{tabular}{|l|c|c|}
\hline
Experiment & Hadron identification technique & Status \\ \hline
CDF & $dE/dx$ in COT & ready for RunII \\ 
& TOF & design, pending funding \\ 
D0 & - &  \\ 
CLEOIII & LiF RICH, chambers with TEA \cite{viehhauser,LiF} & being installed
\\ 
BaBar & DIRC \cite{hocker,dirc} & working \\ 
& $dE/dx$ in DCH & working \\ 
BELLE & Aerogel & working \\ 
& TOF & working \\ 
& $dE/dx$ in CDC & working \\ 
HERA-B & C$_4$F$_{10}$ RICH, PMTs \cite{pyrlik} & working \\ \hline\hline
LHCb & C$_4$F$_{10}$ and aerogel RICH1 & design \\ 
& CF$_4$ RICH2 &  \\ 
& readout with HPDs or PMTs \cite{go} &  \\ 
BTeV & C$_4$F$_{10}$ and aerogel RICH & design \\ 
ATLAS & $dE/dx$ in TRT & design, pending \\ 
&  & beamtest verification \\ 
CMS & - &  \\ \hline
\end{tabular}
\caption{Charged hadron identification techniques in different experiments.}
\label{hid}
\end{table}

The HERA-B RICH detector has proven that a Cherenkov counter can be operated
successfully in the harsh forward hadron spectrometer environment 
\cite{pyrlik}. The RICH has been designed to separate pions and kaons in the
momentum range between 10 and 75 GeV. The radiator used is C$_{4}$F$_{10}$
gas. The photons are imaged by two spherical mirrors and two planar mirrors
onto two photon detection systems outside the spectrometer active volume.
Due to the very high interaction rate and large multiplicity of charged
particles, the Cherenkov photon flux is up to several MHz per cm$^{2}$, and
only photomultipliers were found to operate reliably over the lifetime of
the experiment \cite{krizan}. The photon detectors thus consist of
multianode PMTs, with 16 anodes for the inner region and 4 anodes for the
outer region to match the expected occupancy. In addition, there is a
two-lens telescope in front of each PMT to solve the PMT packing problem
and to match the anode dimension to the dispersion error.

\begin{figure}[tbp]
\mbox{\epsfig{figure=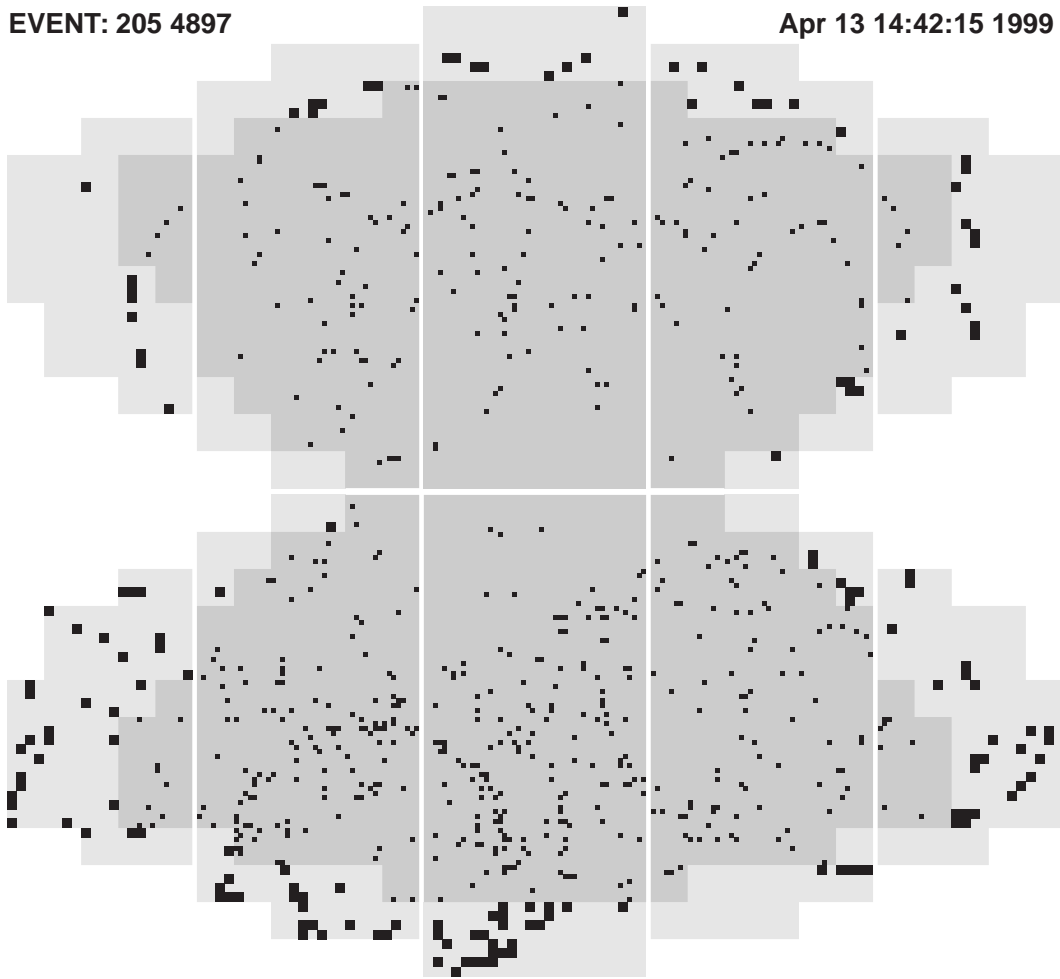,width=14cm}}
\caption{Raw hits of a typical event in the HERA-B RICH 
\protect\cite{pyrlik}. }
\label{herabring}
\end{figure}

The detector was completed in January 1999, and the detector was run under
realistic data-taking conditions in May 1999. Raw hits of a typical event
are shown in Fig.~\ref{herabring}. The rings could be reconstructed in
hadronic events. Even without the final tracking, but using a crude momentum
estimation by matching the calorimeter clusters with the track angles
determined from the RICH and assuming that the particles originate from the
target, the correlation between the Cherenkov ring opening angle and the
particle momentum for different particle species could be demonstrated.

The BaBar experiment has constructed a particle identification system based
on the Detection of Internally Reflected Cherenkov (DIRC) light \cite{dirc}.
The DIRC is a novel concept which uses long, thin, rectangular fused silica
bars both as radiators and as light guides to direct the produced light via
internal reflection to photodetectors. The design has been constrained by
the desired momentum range (700 MeV to 4 GeV), and by the demand of
minimising material in front of the electromagnetic CsI calorimeter. In
addition, for high luminosity running conditions, the particle
identification system must have a fast signal response and be able to
tolerate high backgrounds. These boundary conditions are well satisfied by
the thin fused silica detector elements (17.25 mm thick). Moreover, the
photon detectors can be located outside of the active detector volume. At
one end of the detector bars, the Cherenkov image is allowed to expand
through a standoff region (Standoff Box) filled with water, which has an
index of refraction close to that of fused silica, which minimises the total
reflection at the interface of the two materials. The expanded image is then
detected in an array of PMTs. The instrumented end is located at the
backward direction, since in the asymmetric collisions the final state
particles emerge mainly to the forward direction. The dummy ends of the
detector bars are covered with mirrors to reflect the photons back to the
instrumented end.

\begin{figure}[tbp]
\mbox{\epsfig{figure=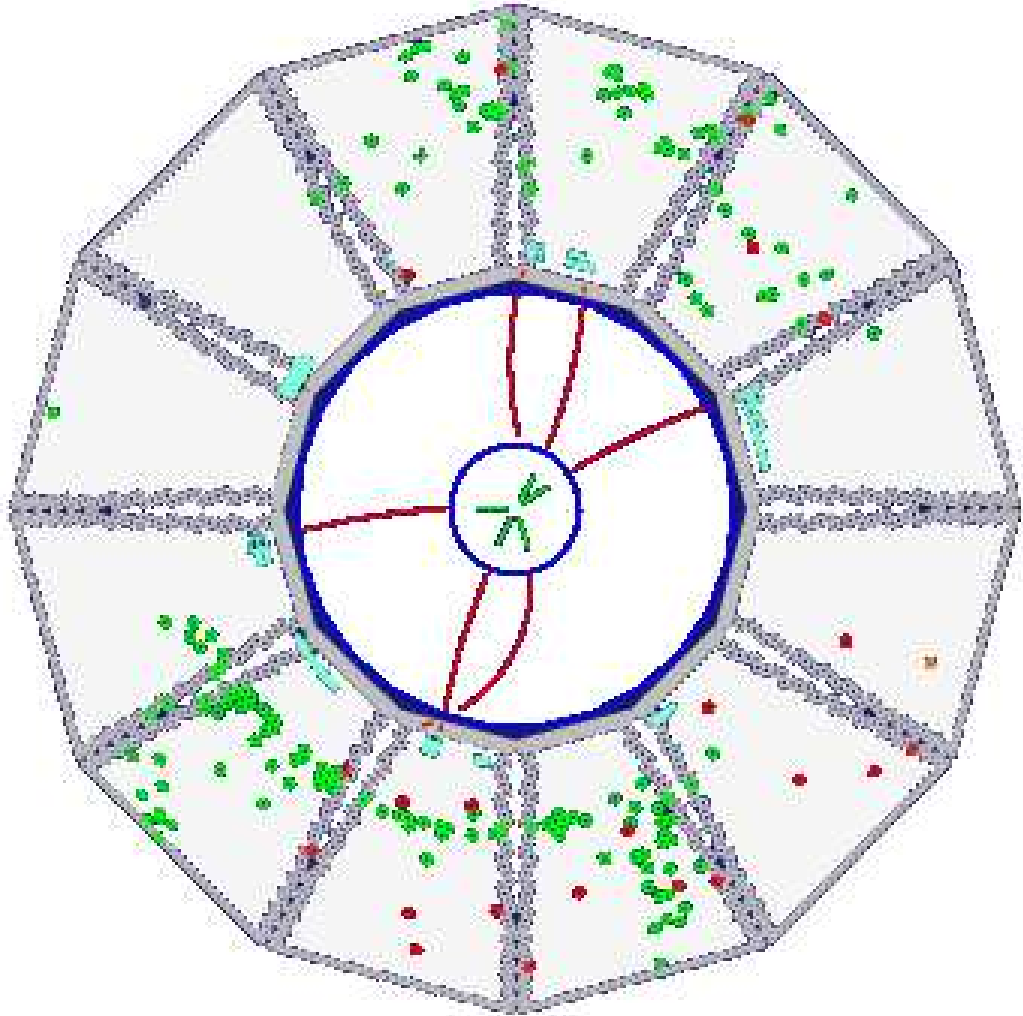,width=14cm}}
\caption{Event display of a multi-hadron event the in the BaBar DIRC 
\protect\cite{hocker}.}
\label{babarring}
\end{figure}

At the time of the conference, five detector elements out of the total 12
were installed in BaBar \cite{hocker}. Cosmic ray data showed a large number
of photons collected and a close agreement of the data with Monte Carlo
predictions. A typical multihadron event reconstructed in the DIRC is shown
in Fig.~\ref{babarring}. The dominant class of background photons originates
from the event itself, while beam-related background can be rejected
effectively by the good time-resolution of the PMTs. The remainder of the
detectors will be installed by end of 1999.

The LHCb experiment has the most difficult task for designing their particle
identification system due to the largest momentum range to be covered, from
1 to 150 GeV. This will be accomplished by using two RICH counters, RICH1
with an aerogel and C$_{4}$F$_{10}$ radiators, and a RICH2 with a CF$_{4}$
radiator \cite{go}. At LHCb, however, there is a strong correlation between
the particle polar angle and momentum, and therefore the angular coverages
of the detectors do not have to be the same. The RICH1 is intended for low
momentum tracks and will have 25-300 mrad polar angle coverage. Tilted
spherical mirrors will focus the Cherenkov rings to photodetectors outside
the detector acceptance. The RICH2 is intended for high momentum, and
consequently small polar angle tracks, and it will cover polar angles 10-120
mrad. The Cherenkov photons are reflected with flat mirrors onto spherical
mirrors, which will focus the photons to photodetectors. Three photodetector
options are presently under study: commercial multianode PMTs, pad HPDs, and
pixel HPDs.

\section{Performance comparison of B-physics experiments}

The performance of several B-physics and general purpose experiments is
compared in Table~\ref{comparison}. The ${\rm e^{+}e^{-}}$ B-factory
experiments will collect data in clean conditions, but the performance is
limited by statistics and the fact that only 
${\rm B}_{{\rm d}}^{{\rm 0}}/{\rm B}^{{\rm +}}$ 
mesons can be produced. Presenting the BaBar estimates as
an example, and assuming 30 fb$^{-1}$ integrated luminosity, $\sin 2\beta $
will be measured with an accuracy of 0.12 using the 
${\rm B_{d}^{0}\rightarrow J/\psi K_{S}^{0}}$ decays (charged particle final
states) -- the precision will be improved to about 0.08 combining various
other modes. The number of reconstructed, untagged events in the 
${\rm B_{d}^{0}\rightarrow \pi ^{+}\pi ^{-}}$ final state will be only 64,
assuming a branching ratio of $4.7\cdot 10^{-6}$. Several years of
data-taking, and combination of other decay modes will be necessary to
obtain reliable information on the angle $\alpha $. The angle $\gamma $
could be studied with the decay mode ${\rm B_{d}^{0}\rightarrow D^{(*)}\pi }$
-- the statistical accuracy of $\sin (2\beta +\gamma )$ would be 0.22. Using
decays ${\rm B^{\pm }\rightarrow DK^{\pm }}$ and 
${\rm B}_{{\rm d}}^{{\rm 0}}{\rm \rightarrow D}^{{\rm 0}}{\rm K^{*}}$, 
300 fb$^{-1}$ would be needed to
obtain a precision in the range 10$^{\circ }$-20$^{\circ }$. Other possible
measurements will be measurements on $|V_{ub}|$ and $|V_{cb}|$, as well as
rare decays of the type ${\rm B\rightarrow K\gamma }$. The BELLE performance
should be similar to BaBar, and if KEKB will be successful in achieving the
design luminosity of 1.0$\cdot $10$^{34}$ cm$^{-2}$s$^{-1}$, the BELLE
experiment should gain about a factor of three in statistics compared to
BaBar.

\begin{table}[htb]
\centering
\begin{tabular}{|l|c|c|c|c|c|}
\hline
Measurement & BaBar & HERA-B & CDF & LHCb & ATLAS \\ 
& BELLE &  & (D0) & BTeV & CMS \\ \hline
$\sin2\beta$ & ** & ** & ** & **** & **** \\ 
$\delta(\sin2\beta)$ & 0.08 & 0.12 & 0.08 & 0.011 & 0.012 \\ \hline
$\sin2\alpha$ & * & * & ** & *** & ** \\ 
N($\pi^+\pi^-$) rec. & 64 unt. & 270 unt. & 4700 unt. & 4600 tagged & 4400
tagged \\ \hline
angle $\gamma$ & * &  & * & *** &  \\ 
& ${\rm B_d^0\rightarrow D^{(*)}\pi }$, &  & ${\rm B^\pm \rightarrow 
DK^{\pm}}$, & ${\rm B_d^0\rightarrow D^{(*)}\pi }$, &  \\ 
& ${\rm B^\pm \rightarrow DK^{\pm }}$, &  & ${\rm B_s^0\rightarrow 
D_s^- K^+}$ & ${\rm B_d^0\rightarrow DK^*}$, &  \\ 
& ${\rm B_d^0\rightarrow DK^*}$ &  &  & ${\rm B_s^0\rightarrow D_s^- K^+}$ & 
\\ \hline
${\rm B_s}$ mixing &  & * & ** & **** & *** \\ 
$\Delta m_s$ reach &  & 12 ps$^{-1}$ & 25 ps$^{-1}$ & 51 ps$^{-1}$ & 
40 ps$^{-1}$ \\ \hline
${\rm B_s}$ analyses &  & * & ** & **** & *** \\ 
$\delta(\delta \gamma))$ from &  &  &  &  &  \\ 
${\rm B_s \rightarrow J/\psi \phi}$ &  &  & 0.16 & 0.01 & 0.03 \\ \hline
${\rm B\rightarrow \mu\mu(X)}$ &  &  &  & *** & *** \\ \hline\hline
Integrated ${\cal L}$& 30 fb$^{-1}$ & 30 fb$^{-1}$ & 2 fb$^{-1}$ & 
2 fb$^{-1}$ & 30 fb$^{-1}$ \\ \hline
\end{tabular}
\caption{Performance comparison of different experiments. Where applicable,
**** indicates a 1\% measurement, *** a few\% measurement, ** a 10\%
measurement and * a measurement. The branching ratio for the decay 
${\rm B_{d}^{0}\rightarrow \pi ^{+}\pi ^{-}}$ was assumed to be 
$4.7\cdot 10^{-6}$ \protect\cite{jaffe}. 
For the ${\rm B_s \rightarrow J/\psi \phi}$ analysis $x_s=20$ was assumed.}
\label{comparison}
\end{table}

Hadronic experiments will be more challenging in view of the much smaller
signal-to-background ratio. Nevertheless, experience from CDF has already
shown the feasibility of achieving significant results in a hadron collider
experiment. As we gain more insight on the complicated pattern of
CP-violating phenomena, more and more channels will turn out to be important
to be experimentally reachable to constrain theoretical uncertainties and to
control experimental systematic uncertainties. This leads to the following
requirements:

\begin{itemize}
\item large statistics,
\item access to ${\rm B^{\pm }/B_{d}^{0}/B_{s}^{0}/B_{c}/\Lambda }_{{\rm b}}$ 
hadrons,
\item trigger on purely hadronic final states,
\item charged hadron identification.
\end{itemize}

While the HERA-B RICH is proving the feasibility of charged hadron
identification in hadronic experiments, the statistics will be the limiting
factor of the HERA-B physics programme. After one year of data-taking, it
is foreseen that the $\sin 2\beta $ will be measured with an accuracy of
0.13 using the ${\rm B_{d}^{0}\rightarrow J/\psi K_{S}^{0}}$ decays. The
number of reconstructed, untagged events in the ${\rm B_{d}^{0}\rightarrow
\pi ^{+}\pi ^{-}}$ final state will be about 270, assuming a branching ratio
of $4.7\cdot 10^{-6}$. The expected background, however, cannot be reliably
estimated. The ${\rm B_{s}^{0}}$ mixing parameter $\Delta m_{s}$ will be
reachable up to about 12 ps$^{-1}$\cite{abt}.

CDF (and D0) will be serious competitors for the dedicated B-experiments at
the first phase of B-experimentation before the LHC startup. CDF at the
RunII will fulfil all the requirements above, while D0 will be lacking
hadron identification and hadronic trigger. After two years of data-taking
(2 fb$^{-1}$), CDF expects to measure $\sin 2\beta $ with an accuracy of
0.084 (0.07 with a kaon tag using the TOF). The number of reconstructed,
untagged events in the ${\rm B_{d}^{0}\rightarrow \pi ^{+}\pi ^{-}}$ final
state will be about 4700, assuming a branching ratio of $4.7\cdot 10^{-6}$.
A 1.3$\sigma $ ${\rm \pi /K}$ separation is expected for particles with
transverse momenta greater than 2 GeV using the $dE/dx$ in the COT, thus
helping to extract the signal from the two-body background. CDF has studied
possibilities for extracting the angle $\gamma $ by using decays 
${\rm B^{\pm }\rightarrow DK^{\pm }}$ and 
${\rm B}_{{\rm s}}^{{\rm 0}}{\rm 
\rightarrow D}_{{\rm s}}^{{\rm -}}{\rm K^{+}}$, but the preliminary studies
indicate a poor accuracy. The decay mode ${\rm B_{s}^{0}\rightarrow J/\psi
\phi }$ will be experimentally easy, and observing an asymmetry in this
decay mode would signal the existence of an anomalous CP violating phase.
The parameter $\delta \gamma $ can be measured with a precision of 
0.16 ($x_{s}=20$). 
The ${\rm B_{s}^{0}}$ mixing parameter $\Delta m_{s}$ will be
reachable up to about 25 ps$^{-1}$ with the CDF baseline detector (20 000
signal events, signal-to-background ratio of 1:2).

The dedicated B-experiments LHCb at LHC and BTeV at Tevatron are being
designed to fulfil all the requirements above in an optimal way. After one
year of data-taking (2 fb$^{-1}$), the LHCb experiment will have measured 
$\sin 2\beta $ with an accuracy of the order of 1\%. The number of
reconstructed and tagged events in the 
${\rm B_{d}^{0}\rightarrow \pi^{+}\pi ^{-}}$ 
final state will be about 4600, assuming a branching ratio of 
$4.7\cdot 10^{-6}$. Various modes for extracting the angle $\gamma $ are
being investigated -- here it will be mandatory to use the full power of the
charged hadron identification. The 
${\rm B}_{{\rm d}}^{{\rm 0}}{\rm \rightarrow D}^{{\rm 0}}{\rm K^{*}}$ 
modes will yield a precision of order 10$^{\circ }$ 
for the angle $\gamma $. The decay mode 
${\rm B_{d}^{0}\rightarrow D^{(*)}\pi }$ 
will need five years running to get the
precision on $2\beta +\gamma $ down to 4$^{\circ }$. 
The ${\rm B}_{{\rm s}}^{{\rm 0}}{\rm 
\rightarrow D}_{{\rm s}}^{{\rm -}}{\rm K^{+}}$ decay mode will
allow measuring $\gamma -2\delta \gamma $ with a statistical precision
between 6$^{\circ }$ and 13$^{\circ }$ with one year's data, depending on
the numerical values of the parameters involved. The parameter $\delta
\gamma $ can be measured with a precision of 1\% in one year using decays 
${\rm B_{s}^{0}\rightarrow J/\psi \phi }$ and the ${\rm B_{s}^{0}}$ mixing
parameter $\Delta m_{s}$ will be reachable up to about 51 ps$^{-1}$. Rare
decays of the type ${\rm B\rightarrow K}^{*0}{\rm \gamma }$ can be
reconstructed efficiently.

Compared to the LHCb performance, the BTeV experiment has actually some
advantages despite of the apparent handicap of smaller center-of-mass
energy, resulting in quite similar expectations for the physics reach. The
longer bunch-spacing and the longer interaction region allow accepting
multiple interactions. Smaller boost means that the experiment can be made
shorter and further away from the beam, saving from radiation damage. If the
two-arms and the vertex trigger can be realised this is an obvious big
advantage.

The general purpose experiments ATLAS and CMS at LHC will be competitive
with the dedicated experiments in some channels. In addition, if some
charged particle identification can be realised, the spectrum of the
B-physics programme can be made much wider. ATLAS and CMS are expected to
obtain a very good precision for the measurement of $\sin 2\beta $, at the
level of 1\% with 30 fb$^{-1}$. Various parameters of the 
${\rm B_{s}^{0}}$-meson system will be measured 
for example with ${\rm B_{s}^{0}\rightarrow
D_{s}\pi ,D_{s}a_{1}}$, and ${\rm B_{s}^{0}\rightarrow J/\psi \phi }$ final
states \cite{smizanska}. The ${\rm B_{s}^{0}}$ mixing parameter 
$\Delta m_{s}$ will be reachable up to 
about 40 ps$^{-1}$, and the statistical precision
for $\delta \gamma $ will be about 3\% ($x_{s}=20$). There are some hopes
for the angle $\alpha $ measurement -- in ATLAS, a ${\rm \pi /K}$ separation
is expected to be possible at the one-standard-deviation level using the
TRT. The issue is to be verified with beam test data. The number of
reconstructed and tagged events in the 
${\rm B_{d}^{0}\rightarrow \pi^{+}\pi ^{-}}$ 
final state will be about 4400, assuming a branching ratio of 
$4.7\cdot 10^{-6}$. Various other precision measurements such as B-baryon
lifetimes and polarizations will be feasible. Rare decays of the type 
${\rm B\rightarrow \mu \mu (X)}$ are particularly suitable 
for ATLAS and CMS
since they are self-triggering, and the high-luminosity data of LHC can be
used as well. For example, the decay mode 
${\rm B_{s}^{0}\rightarrow \mu }^{+}\mu ^{-}$, 
for which the SM branching fraction is $3.5\cdot 10^{-9}$,
should be observed.

\section{Conclusions}

We are living exciting times. The first direct measurement of an angle of
the Unitarity Triangle has emerged from CDF when the physicists have learned
to use the limited statistics in an optimal way. Meanwhile, the precision
measurements from CLEO, LEP, SLD, CDF and kaon experiments are
constraining more and more the Unitarity Triangle. The first generation of
B-physics experiments is getting off the ground and hoping to get to a
cruising speed soon, while the Tevatron experiments are expected to continue
producing even more exciting results in the coming run. From years of
planning, designing and reviewing experiments we are now getting into the
experimentation itself, and hopefully already in the next BEAUTY conference
new results can be reported.

It is also important to discuss the future developments. Therefore, the
future projects at LHC and at Tevatron have their due place in this
conference. When we learn more about B physics, the experiments can be
better optimised to cover fully the physics spectrum, while new ideas,
technological progress and experience help us to design better particle
detectors.

\section{Acknowledgements}

Thanks to Peter Krizan and other local organizers for a well-organized
conference in a beautiful location, to Peter Schlein and Samim Erhan for
making and keeping up 
the spirit of this conference series, to all the conference
participants for an interesting and enjoyable week, and to all the authors
of the contributed papers for providing me with the material for this
summary.

\end{document}